\documentclass[twoside,11pt]{article}

\usepackage{blindtext}

\usepackage[abbrvbib, preprint]{jmlr2e}

\usepackage{mathtools}

\usepackage{bm}
\usepackage{amsbsy,mathptmx}

\usepackage{lastpage}
\usepackage[cal = cm, scr = dutchcal]{mathalpha}

\usepackage[table]{xcolor}
\definecolor{ao}{rgb}{0.0, 0.5, 0.0}

\usepackage{subcaption}

\usepackage{pifont, upgreek, textgreek, xparse}
\DeclareSymbolFont{Symbols}{OMS}{zplm}{m}{n}
\DeclareMathSymbol{\Infty}{\mathord}{Symbols}{"31}

\usepackage{multicol}
\usepackage{enumitem}
\usepackage{booktabs}
\usepackage{algorithm}
\usepackage{algpseudocode}
\usepackage{caption}
\def\argmax{\mathop{\rm argmax}}

\NewDocumentCommand{\grad}{e{_^}}{%
  \mathop{}\!
  \nabla
  \IfValueT{#1}{_{\!#1}}
  \IfValueT{#2}{^{#2}}
}

\ShortHeadings{Binary Spike-and-Slab Lasso Biclustering}{Sijian Fan, Ray Bai}
\firstpageno{1}

\begin{document}

\title{\textsc{BiSSLB}: Binary Spike-and-Slab Lasso Biclustering}

\author{\name Sijian Fan \email sfan@email.sc.edu \\
       \addr Department of Statistics\\
       University of South Carolina\\
       \name Ray Bai \email rbai2@gmu.edu \\
       \addr Department of Statistics\\
       George Mason University}

\editor{My editor}

\maketitle

\begin{abstract}
\textbf{Motivation} \\
Biclustering is a powerful unsupervised learning technique for simultaneously identifying coherent subsets of rows and columns in a data matrix, thus revealing local patterns that may not be apparent in global analyses. However, most biclustering methods are developed for continuous data and are not applicable for binary datasets such as single-nucleotide polymorphism (SNP) or protein-protein interaction (PPI) data. Existing biclustering algorithms for binary data often struggle to recover biclustering patterns under noise, face scalability issues, and/or bias the final results towards biclusters of a particular size or characteristic. \\ 
\textbf{Results} \\
We propose a Bayesian method for biclustering binary datasets called Binary Spike-and-Slab Lasso Biclustering (BiSSLB). Our method is robust to noise and allows for overlapping biclusters of various sizes without prior knowledge of the noise level or bicluster characteristics. BiSSLB is based on a logistic matrix factorization model with spike-and-slab priors on the latent spaces. We further incorporate an Indian Buffet Process (IBP) prior to automatically determine the number of biclusters from the data.  We develop a novel coordinate ascent algorithm with proximal steps which allows for scalable computation. The performance of our proposed approach is assessed through simulations and two real applications on HapMap SNP and Homo Sapiens PPI data, where BiSSLB is shown to outperform other state-of-the-art binary biclustering methods when the data is very noisy.

\end{abstract}

\begin{keywords}
  binary datasets, spike-and-slab lasso, biclustering, matrix factorization, latent factor representation
\end{keywords}

\section{Introduction}\label{sec:intro}

Binary data are ubiquitous in biology-related domains. For example, gene expression data is often binarized into high (``1'') or low (``1'') expression for downstream analysis \citep{gusenleitner2012ibbig, goodwin2016coming}. Single Nucleotide Polymorphism (SNP) data is often converted from categorical genotypes into binary variables, e.g., mutant (``1'') or wild-type (``0'') alleles  \citep{serre2008correction}. In protein-protein interaction (PPI) analysis, a ``1'' (resp. ``0'') signifies a direct physical interaction (resp. lack of interaction) between two specific proteins that control a cellular process \citep{sriwastava2023rubic}.

In these applications, it is frequently of interest to identify localized, coherent patterns in the data. For example, cancer and many other diseases are heterogeneous, comprised of a number of subtypes. Patients belonging to the same disease subtype often exhibit similar gene expression profiles over only a small subset of genes \citep{moran2021spike, Wang2024survey}. Unlike traditional clustering methods which group samples based on \emph{all} of their features, biclustering simultaneously groups samples and features to unveil new subtypes and other local patterns. In the context of binary data, biclustering aims to group observations and features into coherent submatrices of a binary data matrix based on 0/1 patterns \citep{prelic2006systematic, rodriguez2011biclustering}. 

The majority of biclustering algorithms have been developed for continuous or numeric data \citep{cheng2000biclustering, Ihmels2003, tanay2002discovering, FABIA2010, moran2021spike}. However, these approaches are not suitable for binary data. Thus, several biclustering methods specifically tailored for binary data have been developed. \citet{prelic2006systematic} introduced the Bimax algorithm, which uses divide-and-conquer to recursively partition the binary data matrix into smaller submatrices of 1's. \citet{rodriguez2011biclustering} later introduced the Bibit method, which converts binary data into bit words (or integer representations) to efficiently find maximal biclusters. Unfortunately, both Bimax and Bibit are highly sensitive to noise, or the presence of 0's within 1-filled patterns. Since Bimax and Bibit do not allow 0's in the biclusters, they may erroneously estimate a very large number of small biclusters when there is noise in the data \citep{gusenleitner2012ibbig}. This in turn can lead to very high memory and processing time requirements, making them less scalable for large, noisy datasets \citep{LopezFernandez2025}.   

To better handle noisy datasets, probabilistic approaches to binary biclustering have been introduced. Unlike Bimax and Bibit, these models explicitly model the uncertainty in the data, making them more robust to noise. \citet{gusenleitner2012ibbig} proposed the Iterative Binary Bi-clustering of Gene sets (iBBiG) method, which uses a genetic algorithm to maximize the tradeoff between cluster homogeneity (i.e., the number of 1's) and cluster size. iBBiG further uses a fitness score to evaluate the probability of a bicluster being statistically significant. Recently,  \citet{li2020bayesian} introduced Generalized Biclustering (GBC), a Bayesian approach for biclustering of discrete or mixed-type data based on a latent sparse factor model.

The aforementioned methods have several limitations. In addition to issues with scalability and sensitivity to noise, both Bimax and Bibit require specifying the minimum number of rows and columns allowed in a bicluster. iBBiG does not require prior knowledge of the number or size of biclusters and is more scalable. However, the results from iBBiG are highly sensitive to the choice of a modulation parameter $\alpha$ which determines the trade-off between cluster homogeneity and size \citep{gusenleitner2012ibbig}. A small $\alpha$ results in smaller, highly homogeneous clusters, whereas a large $\alpha$ results in larger, more heterogeneous clusters with a greater proportion of 0's mixed with 1's \citep{gusenleitner2012ibbig}. Thus, iBBiG may find biclusters that are either all small or all large. GBC is also more robust to noise than Bimax or Bibit. However, GBC is implemented using an Expectation-Maximization (EM) algorithm, which can suffer from slow convergence \citep{Liu1998PXEM}. In all our experiments and real data applications, we were unable to fit GBC with more than 50 columns in the latent space.

To simultaneously overcome issues with sensitivity to noise, bias towards certain bicluster sizes or counts, and scalability, we propose Binary Spike-and-Slab Lasso Biclustering (BiSSLB), a new Bayesian method for biclustering binary datasets. Our approach extends the Spike-and-Slab Lasso (SSLB) method of \citet{moran2021spike} to the setting of noisy binary data matrices. The original SSLB model in \citet{moran2021spike} assumed continuous data with Gaussian noise, making SSLB unsuitable for binary data. 

BiSSLB utilizes a logistic matrix factorization model with spike-and-slab priors on the latent factor matrices. This sparse latent factor model allows the estimated biclusters to overlap and to be of varying (both small and large) sizes \citep{moran2021spike}. To avoid needing to fix the number of biclusters, we employ an Indian Buffet Process (IBP) prior \citep{ghahramani2005infinite} which automatically learns the number of biclusters from the data. Departing from \citet{moran2021spike}, we also introduce a scalable coordinate ascent algorithm with proximal steps for implementation of BiSSLB. This is in contrast to the original SSLB model, which utilized an EM algorithm. Through simulation studies and applications to real SNP and PPI data, we demonstrate that BiSSLB outperforms other binary biclustering methods, especially in the presence of moderate or high levels of noise.

\section{Materials and methods}\label{sec:method}

\subsection{Logistic matrix factorization}

Let $\mathbf{Y}$ be an $I \times J$ binary data matrix. We assume a logistic matrix factorization model \citep{chen2023statistical} where each $(i,j)$th entry $y_{ij} \in \{ 0, 1 \}$ in $\mathbf{Y}$ independently follows a Bernoulli($p_{ij}$) distribution, and the success probability $p_{ij} = P(Y_{ij}=1)$ is linked to the $(i,j)$th entry $m_{ij}$ of a latent matrix $\mathbf{M} \in \mathbb{R}^{I \times J}$ through
\begin{align} \label{logistic-model}
    p_{ij} = \frac{\exp(m_{ij})}{1+\exp(m_{ij})},
\end{align}
where
\begin{equation} \label{shifted-latent-space}
    \mathbf{M}= \boldsymbol{\mu}\mathbf{1}^{\top} + \mathbf{AB}^{\top}. 
\end{equation}
In \eqref{shifted-latent-space}, $\mathbf{A} \in \mathbb{R}^{I \times K}$ and  $\mathbf{B} \in \mathbb{R}^{J \times K}$ are factor matrices with latent dimension $K$, $\boldsymbol{\mu} \in \mathbb{R}^{I}$ is a vector of location parameters, and $\mathbf{1} \in \mathbb{R}^{J}$ is a vector of all ones. The addition of $\boldsymbol{\mu}$ in \eqref{shifted-latent-space} allows us to capture additional heterogeneity in the samples by modifying the baseline log-odds that $y_{ij} = 1$ \citep{lee2014biclustering, li2020bayesian}.
Under \eqref{logistic-model}-\eqref{shifted-latent-space}, the likelihood function is
\begin{equation} \label{BiSSLB-likelihood}
    p(  \mathbf{Y} \mid \mathbf{A}, \mathbf{B}, \boldsymbol{\mu} ) = \prod_{i=1}^I \prod_{j=1}^J p_{i j}^{y_{i j}}\left(1-p_{i j}\right)^{1-y_{ij}}.
\end{equation}
Since BiSSLB models the underlying probabilities of the entries in $\mathbf{Y}$ being 1's, it is able to handle the inherent uncertainty in noisy binary datasets.

\subsection{Binary spike-and-slab lasso biclustering model}

If both factor matrices $\mathbf{A}$ and $\mathbf{B}$ in \eqref{shifted-latent-space} are sparse, then the biclusters will manifest as nonzero (and possibly overlapping) rank one submatrices in the matrix product $\mathbf{A}\mathbf{B}^\top$ \citep{moran2021spike}. In this case, $K$ (i.e., the number of columns in $\mathbf{A}$ and $\mathbf{B}$) will represent the number of biclusters. Thus, to unveil subgroups of rows and columns, we adopt a Bayesian approach with sparsity-inducing priors on $\mathbf{A}$ and $\mathbf{B}$ in \eqref{shifted-latent-space}. Let $a_{ik}$ and $b_{jk}$ denote the $(i,k)$th and $(j,k)$th entries of $\mathbf{A}$ and $\mathbf{B}$ respectively. We endow these entries with spike-and-slab lasso (SSL) priors \citep{rovckova2018spike}, 
\begin{equation} \label{SSL-priors}
    \begin{aligned}
        &\pi(a_{ik} \mid \widetilde{\gamma}_{ik})=\left(1-\widetilde\gamma_{ik}\right) \psi \left(a_{ik} \mid \widetilde{\lambda}_0 \right)+\widetilde\gamma_{ik} \psi \left(a_{ik} \mid \widetilde{\lambda}_1 \right), \\
        &\pi(b_{jk} \mid \gamma_{jk})=\left(1-\gamma_{jk}\right) \psi \left(b_{jk} \mid \lambda_0 \right)+\gamma_{jk}\psi\left(b_{jk} \mid \lambda_1 \right),
    \end{aligned}
\end{equation}
where $\psi(x \mid \lambda ) = (\lambda / 2)\exp\left(-\lambda | x |\right)$ denotes a Laplace density with inverse scale parameter $\lambda$. In \eqref{SSL-priors}, we set $\widetilde{\lambda}_0 \gg \widetilde{\lambda}_1$ and $\lambda_0 \gg \lambda_1$ so that the mixture components $\psi(\cdot \mid \widetilde{\lambda}_0)$ and $\psi(\cdot \mid \lambda_0)$, or the ``spikes,'' are heavily concentrated around zero. Meanwhile, the other mixture components $\psi(\cdot \mid \widetilde{\lambda}_0)$ and $\psi(\cdot \mid \lambda_1)$, or the ``slabs,'' are relatively diffuse with a large variance. The spikes captures sparsity in $\mathbf{A}$ and $\mathbf{B}$, while the slabs model the nonzero entries. Under \eqref{SSL-priors}, $\widetilde{\gamma}_{ik} \in \{0,1\}$ and $\gamma_{jk} \in \{0,1\}$ are binary variables indicating whether the corresponding entry $a_{ik}$ or $b_{jk}$ is drawn from the spike or the slab component. 

Given the indicator variables $\widetilde{\gamma}_{ik}$ and $\gamma_{jk}$ in \eqref{SSL-priors}, we define the binary matrices $\boldsymbol{\widetilde{\Gamma}} = (\widetilde{\gamma}_{ik})$ and $\boldsymbol{\Gamma} = (\gamma_{jk})$ which respectively have $I$ and $J$ rows but potentially an infinite number of columns. To adaptively learn the effective column dimension $K$ in $\mathbf{A}$ and $\mathbf{B}$, we follow \cite{moran2021spike} and couple the SSL priors \eqref{SSL-priors} with Indian buffet process (IBP) priors \citep{ghahramani2005infinite} on $\widetilde{\boldsymbol{\Gamma}}$ and $\boldsymbol{\Gamma}$, 
\begin{equation} \label{IBP-priors}
    \begin{aligned}
        &\widetilde{\gamma}_{ik} \mid \widetilde{\theta}_{(k)} \sim \text{Bernoulli}(\widetilde{\theta}_{(k)}), & \widetilde{\theta}_{(k)} = \prod_{l=1}^{k} \widetilde{\nu}_{l}, &~~~ \widetilde{\nu}_l \overset{\text{iid}}{\sim} \text{Beta}(\widetilde{\alpha}, 1), & k = 1, 2, \ldots, \\
        &\gamma_{jk} \mid \theta_{(k)} \sim \text{Bernoulli}(\theta_{(k)}), &  \theta_{(k)} = \prod_{l=1}^{k} \nu_{l}, &~~~ \nu_l \overset{\text{iid}}{\sim} \text{Beta}(\alpha, 1), & k = 1, 2, \ldots,  
    \end{aligned}
\end{equation}
where $\widetilde{\alpha} > 0$ and $\alpha > 0$ are intensity parameters. The priors \eqref{IBP-priors} are based on the stick-breaking representation of the IBP \citep{teh2007stick}. It can be shown that the Bernoulli probablities $\widetilde{\theta}_{(k)}$ and $\theta_{(k)}$ in \eqref{IBP-priors} are exponentially decreasing with $k$ \citep{teh2007stick}. This induces a left-ordered structure where, with overwhelming probability, the leftmost columns of $\boldsymbol{\widetilde{\Gamma}}$ and $\boldsymbol{\Gamma}$ are dense with many nonzero entries, while the later columns are all zero vectors. This allows us to adaptively learn the number of biclusters $K$, since the rightmost columns of $\boldsymbol{\widetilde{\Gamma}}$ and $\boldsymbol{\Gamma}$ are not activated \citep{rovckova2016fast, moran2021spike}. 

The SSL-IBP formulation \eqref{SSL-priors}-\eqref{IBP-priors} also induces a soft identifiability constraint on our model \citep{rovckova2016fast, moran2021spike}. Without any constraints, $\mathbf{A}$ and $\mathbf{B}$ in \eqref{shifted-latent-space} are not identifiable since we can always post-multiply them by an orthogonal matrix $\mathbf{P}$ and we would get an identical likelihood function \eqref{BiSSLB-likelihood}, i.e., $(\mathbf{AP})(\mathbf{BP})^\top = \mathbf{A}\mathbf{B}^\top$. 
By anchoring our priors \eqref{SSL-priors}-\eqref{IBP-priors} on a sparse factorization of $\mathbf{AB}^\top$, we are more likely to orient our model towards an identifiable, sparse configuration of $\mathbf{AB}^\top$ without needing to impose hard constraints such as requiring that the columns of $\mathbf{A}$ and $\mathbf{B}$ be orthonormal or that $\mathbf{B}$ be lower triangular \citep{rovckova2016fast, moran2021spike}. 

To complete our prior specification, we place a flat noninformative prior on the location vector $\boldsymbol{\mu}$ in \eqref{BiSSLB-likelihood}, i.e.,
\begin{equation} \label{mu-prior}
    \pi(\boldsymbol{\mu}) \propto 1. 
\end{equation}

\subsection{Coordinate ascent algorithm}\label{sec:algorithm}

We aim to find the posterior mode $(\widehat{\mathbf{A}}, \widehat{\mathbf{B}}, \widehat{\boldsymbol{\mu}})$ for the parameters under the BiSSLB model \eqref{logistic-model}-\eqref{mu-prior}. Since the modes for $\mathbf{A}$ and $\mathbf{B}$ are exactly sparse \citep{rovckova2018spike, rovckova2016fast}, we can then identify the biclusters from the matrix product $\widehat{\mathbf{A}} \widehat{\mathbf{B}}^\top$. 

Spike-and-slab models, including those for factor analysis \citep{rovckova2016fast, moran2021spike}, are commonly implemented using the EM algorithm where the binary indicator variables (e.g. $\widetilde{\gamma}_{ik}$ and $\gamma_{jk}$ in \eqref{SSL-priors}) are marginalized out in the E-step and then the subsequent parameters are maximized in the M-step \citep{rockova2014emvs,Tang2017genetics,bai2026bayesian}. However, the EM algorithm can suffer from slow convergence and/or convergence to suboptimal local maxima \citep{Liu1998PXEM,rovckova2016fast}. Although techniques such as parameter expansion or deterministic annealing can partially mitigate these problems \citep{Liu1998PXEM,rovckova2016fast, rockova2014emvs}, we instead introduce a coordinate ascent algorithm with proximal gradient steps \citep{neal2014proximal}. Our method bypasses the EM algorithm entirely, thus avoiding the issues associated with the EM algorithm.

The sharp exponential decay of the $\widetilde{\theta}_{(k)}$'s and $\theta_{(k)}$'s to zero in \eqref{IBP-priors} suggests that in practice, we can use a truncated approximation to the IBP by setting $\theta_{(k)}=0$ for all $k > K^{\star}$, where $K^{\star}$ is an overestimate of the number of biclusters \citep{rovckova2016fast, moran2021spike}. We adopt this truncation strategy for the practical implementation of BiSSLB.

  Let $\pi(\mathbf{a}_k)$ and $\pi(\mathbf{b}_k)$ denote the marginal prior distributions for the $k$th columns of $\mathbf{A}$ and $\mathbf{B}$ respectively after integrating out $(\widetilde{\boldsymbol{\Gamma}}, \widetilde{\boldsymbol{\theta}})$ and $(\boldsymbol{\Gamma}, \boldsymbol{\theta})$ in the hierarchical priors \eqref{SSL-priors}-\eqref{IBP-priors}. The log-posterior under BiSSLB is then
\begin{equation} \label{BiSSLB-objective}
    \begin{aligned}
        \mathcal{L} \left( \mathbf{A}, \mathbf{B}, \boldsymbol{\mu} \right) =
        & 
        \sum_{i=1}^{I} \sum_{j=1}^{J} \left[\mathbf{Y} \odot\left(\boldsymbol{\mu}\mathbf{1}^\top+\mathbf{AB^{\top}}\right)-\log\left(1+\exp\left(\boldsymbol{\mu}\mathbf{1}^\top+\mathbf{AB^{\top}}\right)\right)\right]_{i j}
        +\sum_{k=1}^{K^{\star}} \left[ \log \pi(\mathbf{a}_k) + \log \pi(\mathbf{b}_k) \right], 
    \end{aligned}
\end{equation}
where $\odot$ denotes the Hadamard product, and $[\mathbf{X}]_{ij}$ denotes the $(i,j)$th entry of a matrix $\mathbf{X}$.  

After initializing the parameters, our coordinate ascent algorithm updates each parameter while holding all other parameters fixed. In particular, the updates for $\mathbf{A}$ and $\mathbf{B}$ are
\begin{equation*}
    \mathbf{A}^{(t)} = \argmax_{\mathbf{A} \in \mathbb{R}^{I \times K^{\star}}} \mathcal{L} \left( \mathbf{A}, \mathbf{B}^{(t-1)}, \boldsymbol{\mu}^{(t-1)} \right)  \qquad \text{and} \qquad 
    \mathbf{B}^{(t)} = \argmax_{\mathbf{B} \in \mathbb{R}^{J \times K^{\star}}} \mathcal{L} \left( \mathbf{A}^{(t)}, \mathbf{B}, \boldsymbol{\mu}^{(t-1)} \right),
\end{equation*}
where $\mathcal{L}$ is the BiSSLB log-posterior \eqref{BiSSLB-objective}. The specific updates for $\mathbf{A}^{(t)}$ and $\mathbf{B}^{(t)}$ are derived in Section A.1 of the Supplementary Material and are available in closed form (Equations (A.14)-(A.15) and (A.16)-(A.17) of the Supplement). Briefly, we use a proximal gradient step to manage the non-differentiable terms $\log \pi(\mathbf{a}_k)$ and $\log \pi(\mathbf{b}_k)$ in \eqref{BiSSLB-objective}.  To further accelerate convergence, we incorporate a momentum-based scheme with a learning rate $\eta$, akin to the fast iterative shrinkage-thresholding algorithm (FISTA) proposed by \cite{beck2009fista}. It is worth pointing out that the updates for $\mathbf{A}^{(t)}$ and $\mathbf{B}^{(t)}$ are based on a refined characterization of the global posterior mode for $\mathbf{A}$ and $\mathbf{B}$ \citep{rovckova2018spike}. This enables us to discard many suboptimal local modes for $\mathbf{A}$ and $\mathbf{B}$ in \eqref{BiSSLB-objective}, thus increasing the likelihood that our algorithm converges to a local mode that closely approximates or coincides with the global mode. This also helps to mitigate the issue of local entrapment, a common issue with the EM algorithm \citep{rockova2014emvs, rovckova2016fast}. 

The update for $\boldsymbol{\mu}$ is the maximizer of \eqref{BiSSLB-objective} with respect to $\boldsymbol{\mu}$ (holding $\mathbf{A}$ and $\mathbf{B}$ fixed) and has a closed form (Step 3 of Algorithm \ref{algorithm:bisslb}). 
Finally, our algorithm requires us to update the conditional expectations $\widetilde{\tau}_k = \mathbb{E}[ \widetilde{\theta}_{(k)} \mid \mathbf{a}_k ]$ and $\tau_k = \mathbb{E}[ \theta_{(k)} \mid \mathbf{b}_k ]$ for all $k = 1, \ldots, K^{\star}$, in each iteration  (see Appendix A for more details).  The updates for  $\widetilde{\boldsymbol{\tau}} = (\widetilde{\tau}_1, \ldots, \widetilde{\tau}_{K^{\star}})^\top$ and $\boldsymbol{\tau} = (\tau_1, \ldots, \tau_{K^{\star}})^\top$ also have a closed form (Steps 4-5 of Algorithm \ref{algorithm:bisslb}). These updates are formally derived in Sections A.2 and A.3 of the Supplementary Material. 

The complete BiSSLB coordinate ascent algorithm is given in Algorithm \ref{algorithm:bisslb}. At the end of each iteration, we reset the pseudo-upper bound on the number of biclusters $K^{\star}$, and we further rescale $\mathbf{A}$ and $\mathbf{B}$ to ensure numerical stability \citep{moran2021spike}. Section A.4 of the Supplementary Material discusses how to initialize the parameters in our algorithm.  Note that because we applied a momentum-based update for $\mathbf{A}$ and $\mathbf{B}$, the first iteration ($t=1$) also uses this initialization before proceeding with cyclical updates of all model parameters in each subsequent $t$th iteration, $t \geq 2$. 

\begin{algorithm}[htb!] 
    \caption{BiSSLB Coordinate Ascent Algorithm}\label{algorithm:bisslb}
    
    \textbf{Input:} Binary matrix $\mathbf{Y}$, hyperparameters $\{ \widetilde{\lambda}_0, \widetilde{\lambda}_1, \lambda_0, \lambda_1,  \widetilde{\alpha}, \widetilde{\beta}, \alpha, \beta,  K^{\star} \}$ and learning rate $\eta$
    
    \textbf{Output:} posterior modes $\widehat{\boldsymbol{\mu}}, \widehat{\mathbf{A}}, \widehat{\mathbf{B}}$

    \begin{algorithmic}
    
        \State Initialize: $\boldsymbol{\Omega}^{(0)} = \left\{\mathbf{A}^{(0)}, \mathbf{B}^{(0)}, \boldsymbol{\mu}^{(0)},  \widetilde{\boldsymbol{\tau}}^{(0)}, \boldsymbol{\tau}^{(0)}\right\}$
        \State Set $\boldsymbol{\Omega}^{(1)} = \boldsymbol{\Omega}^{(0)}$ and initialize counter at $t = 2$
        \State 
        \While{not converged} 
            \State 1. Update $\mathbf{A}^{(t)} \gets \underset{\mathbf{A}\in \mathbb{R}^{I\times K^{\star}}}{\arg \max} \ \mathcal{L} (\mathbf{A}, \mathbf{B}^{(t-1)}, \boldsymbol{\mu}^{(t-1)})$ ~ \slash \slash ~  (A.15) in Supplementary Material
        
            \State 2. Update $\mathbf{B}^{(t)} \gets \underset{\mathbf{B}\in \mathbb{R}^{J\times K^{\star}}}{\arg \max} \ \mathcal{L} (\mathbf{A}^{(t)}, \mathbf{B}, \boldsymbol{\mu}^{(t-1)})$ ~~~~ \slash \slash ~ (A.17) in Supplementary Material 
    
            \State 3. Update $\boldsymbol{\mu}^{(t)}$ as
            $$
        \mu_i^{(t)}  \leftarrow \mu_i^{(t-1)}+ \frac{4}{J} \sum_{j=1}^{J} (y_{ij}-p_{ij} ) \quad \text{for all } i = 1, \ldots, I
            $$
            \qquad ~ where $p_{ij} = 1/ \{ 1+\exp(-\mu_i^{(t-1)} - \langle \mathbf{a}_i^{(t)}, \mathbf{b}_j^{(t)} \rangle ) \}$
            \State 4. For all $k = 1, \ldots, K^{\star}$, update $\widetilde{\tau}_k := \mathbb{E}[ \widetilde{\theta}_{(k)} \mid \mathbf{a}_k]$ and $\tau_k := \mathbb{E}[ \theta_{(k)} \mid \mathbf{b}_k ]$ as
            $$
            \widetilde{\tau}_k^{(t)} \gets \frac{\tilde{\alpha} + \|\mathbf{a}_k^{(t)}\|_0}{\tilde{\alpha} + \tilde{\beta} + I} \quad \text{and} 
            \quad \tau_k^{(t)} \gets \frac{\alpha + \|\mathbf{b}_k^{(t)}\|_0}{\alpha + \beta + J}
            $$
            
            \State 5. Reorder $\widetilde{\boldsymbol{\tau}}^{(t)}$ and $\boldsymbol{\tau}^{(t)}$ in descending order, and reorder corresponding columns in $\mathbf{A}^{(t)}$ and $\mathbf{B}^{(t)}$
            
            \State 6. Remove the zero columns and reset $K^{\star}$

            \State 7. Rescale $\mathbf{A}$ and $\mathbf{B}$ for stability:
            $$c_k \leftarrow \sqrt{\frac{\left\|\mathbf{a}_k^{(t)}\right\|_1}{\left\|\mathbf{b}_k^{(t)}\right\|_1}}, \quad \mathbf{a}_k^{(t)} \leftarrow \frac{1}{c_k} \mathbf{a}_k^{(t)}, \quad \mathbf{b}_k^{(t)} \leftarrow c_k \mathbf{b}_k^{(t)}$$
        
            \State 8. Iterate $t \gets t + 1$
            
        \EndWhile
    \end{algorithmic}
\end{algorithm}

\subsection{Hyperparameter settings and software} \label{sec:hyperparameters}

We fix the slab hyperparameters $(\widetilde{\lambda}_1, \lambda_1)$ in the SSL priors \eqref{SSL-priors} as $\widetilde{\lambda}_1 = \lambda_1 = 1$. Meanwhile, we recommend tuning the spike hyperparameters $(\widetilde{\lambda}_0, \lambda_0)$ from a ladder   \citep{moran2021spike}. In all of our simulations and real data applications, we set $\widetilde{\lambda}_0 = \lambda_0$ and tuned them from the grid $\{ 1, 5, 10, 50, 100, 1000, 10{,}000 \}$. In the IBP priors \eqref{IBP-priors}, we fix the shape hyperparameters $(\widetilde{\alpha}, \alpha, \widetilde{\beta}, \beta)$ as $\widetilde{\alpha} = \alpha =  1/K^{\star}$ and $\widetilde{\beta} = \beta_1 = 1$. The learning rate $\eta$ for the proximal gradient updates is set as $\eta = 10^{-3}$ by default, but can also be tuned from a grid, e.g., $\{ 10^{-3}, 10^{-4}, \ldots, 10^{-8} \}$. Finally, as discussed in Section \ref{sec:algorithm}, the pseudo-upper bound $K^{\star}$ should be set to an initial overestimate of the number of true biclusters. The BiSSLB algorithm then adaptively learns an estimate $\widehat{K}$ for the bicluster count. If $K^{\star}$ has not changed by the end of the algorithm, the practitioner can try a larger initial $K^{\star}$. In our two real data applications in Section \ref{sec:realApp}, BiSSLB was able to estimate both a small number ($\widehat{K} = 3$) and a relatively large number ($\widehat{K} = 311$) of biclusters. 

Our method is implemented in an \textsf{R} package \texttt{BiSSLB} which is publicly available at \url{https://github.com/Sijianf/BiSSLB}.

\section{Simulation analysis}\label{sec:simulation}

We considered two simulation settings designed to assess BiSSLB's ability to (a) estimate the true biclusters, and (b) recover the unknown number of biclusters. We compared BiSSLB to Bibit, Bibit2, Bimax, iBBiG, and GBC. Bibit2 is a modification to the original Bibit algorithm, which adds a noise parameter to allow for noisy biclusters, i.e., biclusters containing both 1's and 0's \citep{rodriguez2011biclustering}. For BiSSLB, we set the hyperparameters according to the strategies described in Section \ref{sec:hyperparameters}. For BiSSLB, iBBiG, and GBC, we used an initial guess of 20 biclusters. For all other hyperparameters in the competing methods, we used the default values suggested by the authors. For each simulation setting, we conducted 50 Monte Carlo simulations and examined the performance of the different methods averaged across the 50 replications. 

We compared the different methods in terms of clustering error (CE) \citep{li2020bayesian}, consensus score (CS) \citep{prelic2006systematic}, and relevance and recovery \citep{FABIA2010, moran2021spike}. These metrics lie between 0 and 1, with a higher score indicating better performance for that specific metric. The exact expressions for these metrics are given in Section B of the Supplementary Material. CE and CS measure the amount of overlap between the estimated biclusters and true biclusters after an optimal matching of clusters. Higher values of CE and CS indicate greater overlap. The main difference between CE and CS is that CE takes bicluster size into consideration and assigns greater weight to larger biclusters \citep{li2020bayesian}. Relevance measures how similar the estimated biclusters are to the true biclusters on average, where similarity is defined by the Jaccard index. Finally, recovery measures how similar the true biclusters are to the estimated biclusters on average. Although relevance indicates whether the found biclusters are correct, it does not reflect whether the rest of the pattern was missed. Meanwhile, recovery captures whether true signals were found, but not whether a lot of noise was also included. Thus, CE and CS are considered more comprehensive measures, balancing both relevance and recovery \citep{li2020bayesian, moran2021spike}.

\subsection{Simulation settings}\label{ssec:simulation-settings}

\textbf{Simulation I: Noisy matrix with blocks of varying sizes.} We set $I = 300$ and $J = 1000$ with $K = 15$ true biclusters. Recall that $\mathbf{a}_k$ and $\mathbf{b}_k$ denote the $k$th columns of $\mathbf{A}$ and $\mathbf{B}$ respectively. For each $k = 1, \ldots, K$, we generated a bicluster by randomly setting several consecutive entries in $\mathbf{a}_k$ and $\mathbf{b}_k$ to be ones and setting all remaining entries to be zero; the outer product $\mathbf{a}_k \mathbf{b}_k^\top$ then contained a rank-one submatrix.
For each $k$, the bicluster row dimension $r_k$ was randomly chosen from $ r_k \in \{ 5, 6, \ldots, 20 \}$, and the column dimension $c_k$  was randomly chosen from $ c_k \in \{ 10, 11, \ldots 50 \}$. Thus, the biclusters in the matrix $\mathbf{M} = (m_{ij})$, where $\mathbf{M} = \mathbf{A} \mathbf{B}^\top$, were of varying sizes $r_k \times c_k$ and were allowed to overlap. We then generated the observed binary matrix $\mathbf{Y} = (y_{ij})$ as $y_{ij} = \mathbb{I} (m_{ij} \neq 0)$. Finally, we added random noise by randomly flipping a fixed proportion of the entries in $\mathbf{Y}$ from ``0'' to ``1'' or from ``1'' to ``0.'' This resulted in both 0's in the clusters of 1's and single isolated entries of 1's surrounded by 0's. We considered noise levels of $\{ 0, 0.01, \ldots, 0.20 \}$.

\begin{figure}[H]
    \centering
    \includegraphics[width=0.75\textwidth]{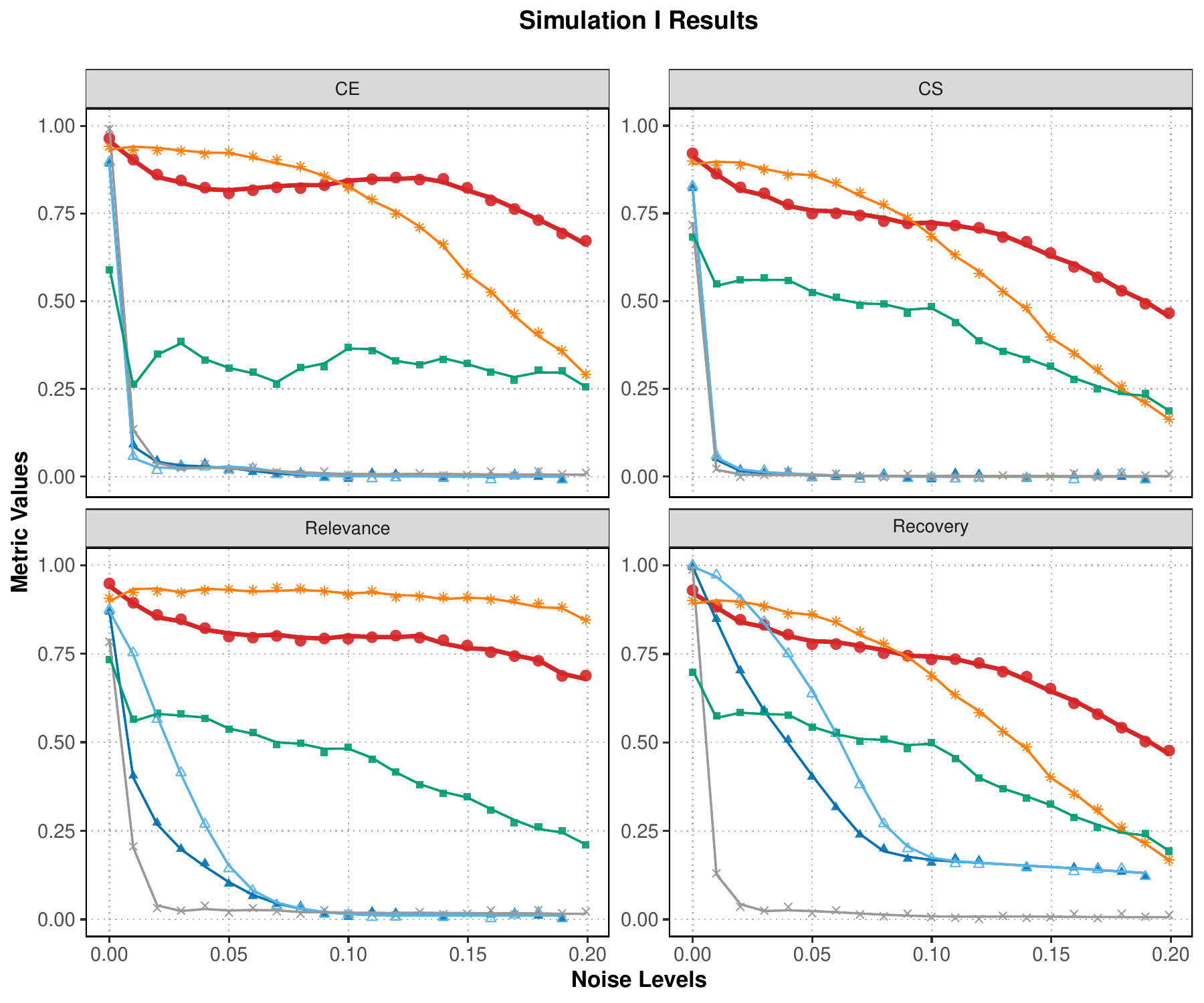}
    \includegraphics[width=0.75\textwidth]{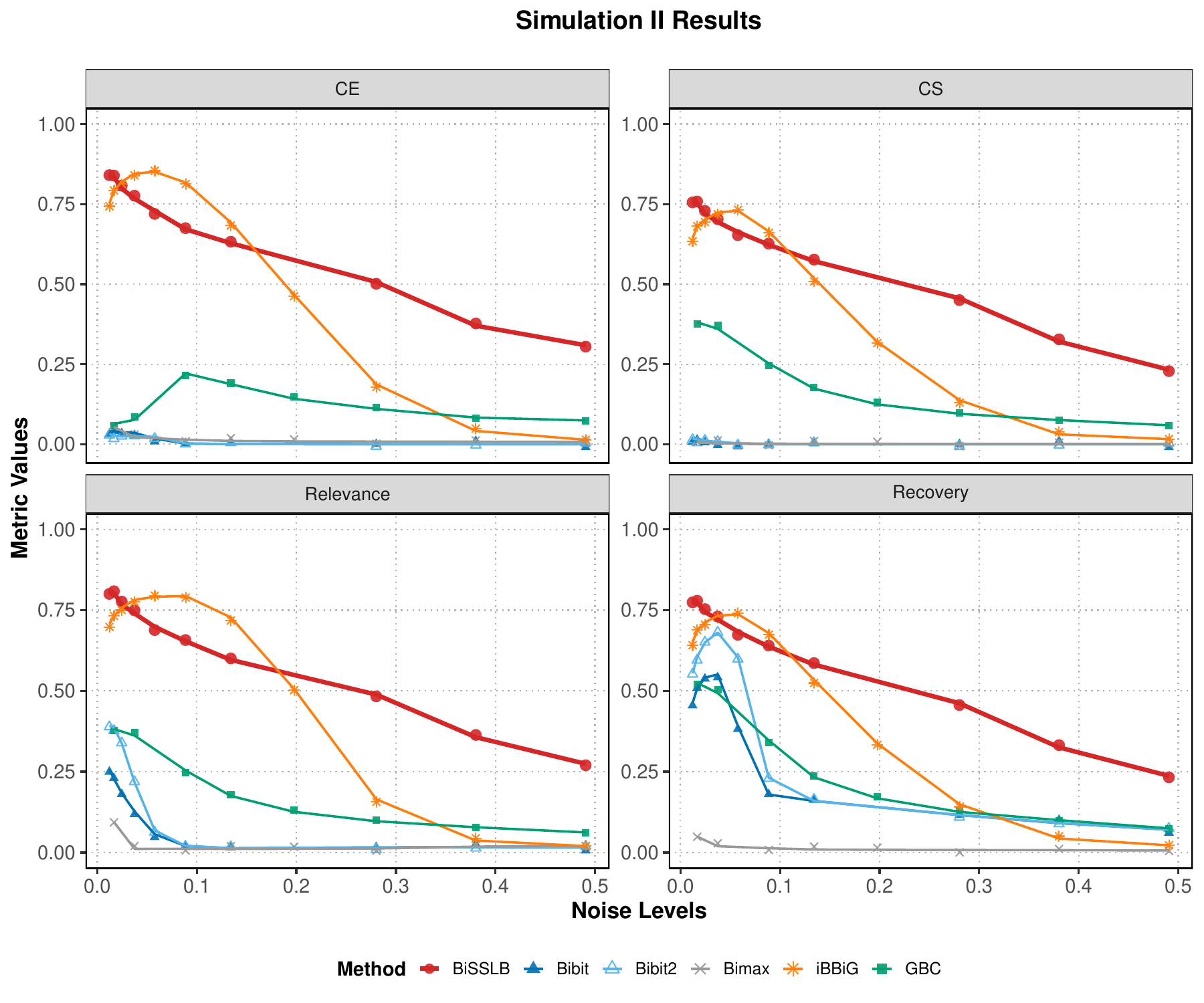}
    \caption{Comparison of the different biclustering algorithms in Simulation I (top four panels) and Simulation II (bottom four panels) in terms of consensus error (CE), consensus score (CS), relevance, and recovery. Each of these metrics is plotted against the noise level. The results displayed were averaged across 50 Monte Carlo replicates.}
    \label{fig:sim_performance_comparison}
\end{figure}

\noindent \textbf{Simulation II: Noisy matrix arising from a generative model.} In the second simulation, we induced the noise in the observed matrix through a generative model. We first generated the matrix product $\mathbf{A}\mathbf{B}^\top$ the same way as in Simulation 1. However, we then assigned random values of $\mathcal{N}(\pm 2, 0.1^2)$ to the entries in the clusters, where the sign of the mean was randomly chosen. The other entries were assigned values from $\mathcal{N}(0, 0.1^2)$. The true data matrix $\mathbf{Y}$ was then generated with entries according to
\begin{equation*}
    y_{ij} \sim \text{Bernoulli} \left( \frac{1}{1+\exp(-(\mu +\mathbf{a}_i^\top \mathbf{b}_j)) } \right),
\end{equation*}
where we varied $\mu \in \{ -5, -4.5, \ldots, 0.5, 0 \}$ to obtain different noise levels. As $\mu$ increases, more noise is added.

\subsection{Simulation results}\label{ssec:simulation-results}

\begin{figure}[t!]
    \centering
    \includegraphics[width=\textwidth]{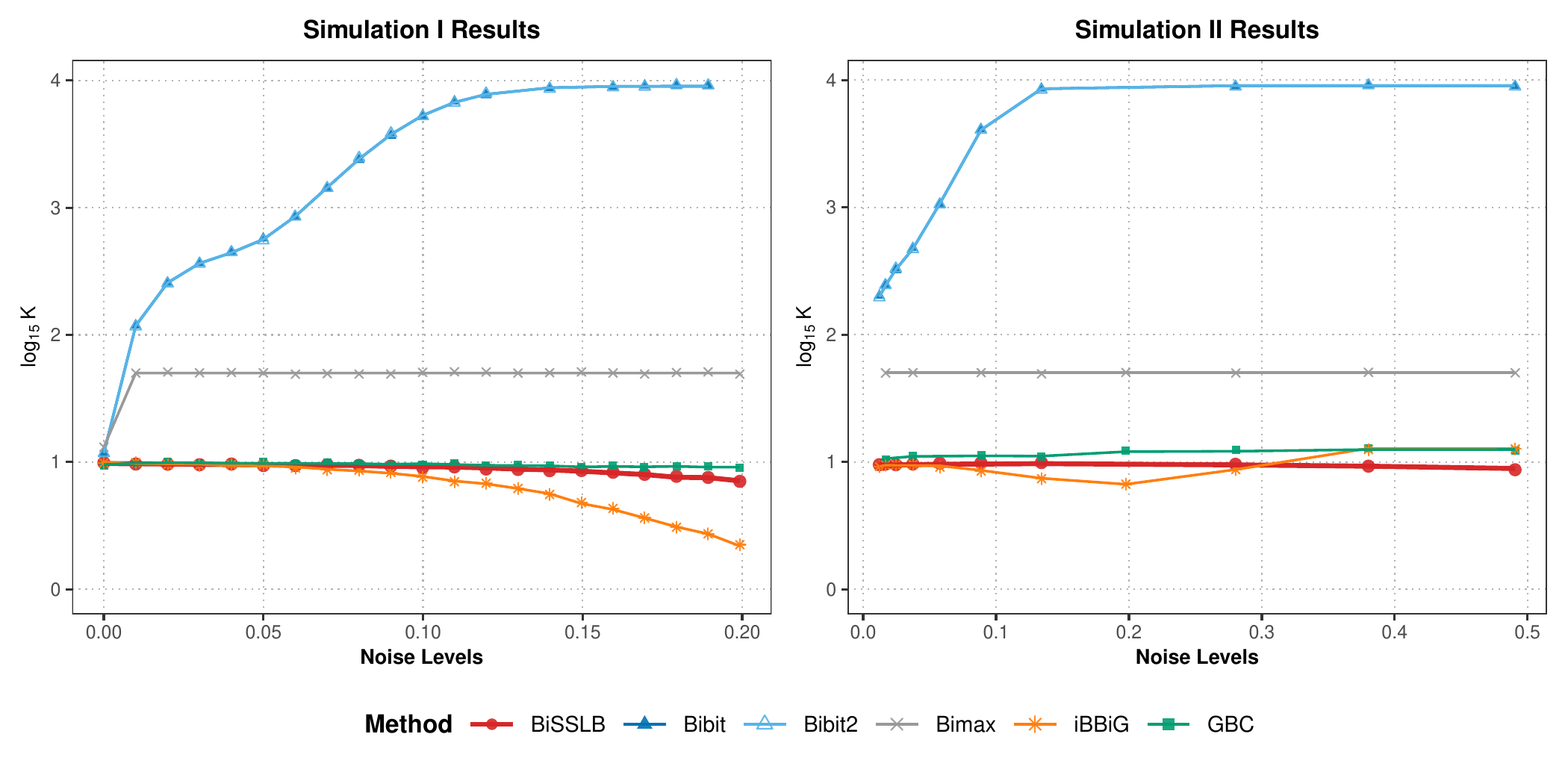}
    \caption{Results for recovery of the true bicluster number ($K=15$) in Simulation I (left panel) and Simulation II (right panel) across different noise levels. In these plots, the vertical axis is $\log_{15}\widehat{K}$, where $\widehat{K}$ is the estimated number of biclusters for each algorithm. Thus, $\log_{15}\widehat{K}=1$ indicates perfect recovery of the number of biclusters. The results displayed were averaged across 50 Monte Carlo replicates.}
    \label{fig:sim_K_comparison}
\end{figure}

Figure \ref{fig:sim_performance_comparison} plots the results for the different methods averaged across 50 replications. We see that BiSSLB and iBBiG were consistently the top two performing methods, with considerably higher average CE, CS, relevance, and recovery than Bibit, Bibit2, Bimax, and GBC once there was non-zero noise in $\mathbf{Y}$. However, once the noise level exceeded 10\%, the performance of iBBiG tended to drop sharply in terms of CS, CE, and recovery in Simulation I (top four panels of Figure \ref{fig:sim_performance_comparison}) and in terms of CS, CE, relevance, and recovery in Simulation II (bottom four panels of Figure \ref{fig:sim_performance_comparison}). BiSSLB significantly outperformed iBBiG in terms of CS, CE, and recovery for all noise levels of 10\% or higher. Simulation II was an especially challenging scenario because the noise level could be as high as 50\%. In Simulation II, all methods unsurprisingly performed worse as the noise level increased; however, BiSSLB still had much higher CE, CS, relevance, and recovery than all other methods for moderate to high noise levels. These results demonstrate that BiSSLB was the most resilient to noise in very noisy binary datasets.

We also assessed the ability of the different methods to recover the true number of biclusters ($K=15$). Let $\widehat{K}$ denote the estimated number of biclusters. Figure \ref{fig:sim_K_comparison} plots $\log_{15} \widehat{K}$ against the noise level for the different biclustering methods, where $\log_{15} \widehat{K}=1$ indicates perfect recovery. We set that Bibit, Bibit2, and Bimax tended to dramatically overestimate the number of biclusters, with $\log_{15} \widehat{K} > 1$ once there was noise introduced in $\mathbf{Y}$.
In Simulation I (left panel of Figure \ref{fig:sim_K_comparison}), iBBiG \emph{underestimated} the true number of biclusters ($\log_{15} \widehat{K} < 1$) once the noise level exceeded 0.10. In contrast, BiSSLB and GBC gave the most consistent performance in Simulation I, with GBC only slightly outperforming BiSSLB for higher noise levels. However, in Simulation II (right panel of Figure \ref{fig:sim_K_comparison}), GBC and iBBiG both \textit{overestimated} the true number of biclusters when the noise levels were moderate or high. Meanwhile, BiSSLB consistently estimated the correct number of biclusters on average in Simulation II, even when the noise level was as high as 50\%. Overall, our simulations show that BiSSLB was the most consistent method for recovering the true number of biclusters under moderate or high levels of noise.

\section{Real applications}\label{sec:realApp}

\subsection{HapMap single nucleotide polymorphism analysis}\label{ssec:snp}

Single nucleotide polymorphism (SNP) genotyping is a widely used technique for identifying variations in DNA sequences among individuals. We analyzed data from the International HapMap Project \citet{international2005haplotype} which was also analyzed by \citet{serre2008correction}. In this dataset, there are 270 samples consisting of three ethnic groups: 90 Caucasians (Utah residents with ancestry from northern and western Europe), 90 Africans (individuals from Yoruba in Ibadan, Nigeria), and 90 Asians (45 Chinese people from Beijing, China and 45 Japanese people from Tokyo, Japan). We chose to analyze these data because the three ground truth clusters (based on ethnicity) were known, allowing us to evaluate the clustering accuracy of different biclustering algorithms.

\cite{serre2008correction} selected 1536 SNPs that were potentially associated with acute myocardial infarction. Among these, we used the 1391 SNPs that were shared among the three ethnic groups after excluding those with missing matches. Each SNP was coded as ``0'' for the most prevalent homozygous genotype (wild-type) and ``1'' for those with an altered DNA sequence  (mutant). This resulted in a $270 \times 1391$ binary matrix.
We implemented BiSSLB and the competing biclustering algorithms (Bibit, Bibit2, Bimax, iBBiG, and GBC) to this data and evaluated clustering accuracy using ancestry information as the ground truth for sample labels. For BiSSLB, iBBiG, and GBC, we set an initial guess of 10 biclusters; we denote the competing methods as iBBiG(10) and GBC(10). All other hyperparameters were set as in Section \ref{sec:simulation}. 

\begin{table}[t!]     
    \centering
    \caption{Comparison of different biclustering methods on the HapMap SNP dataset. For iBBiG and GBC, the numbers in parentheses are the number of biclusters with which these algorithms were initiated. The last column $\widehat{K}$ is the estimated number of biclusters for each method.}
    \begin{tabular}{lccccr}
    \hline
    \textbf{Method} & \textbf{CS} & \textbf{CE} & \textbf{Recovery} & \textbf{Relevance} & $\widehat{\boldsymbol{K}}$ \\ 
    \hline
    BiSSLB    & 0.356 & 0.356 & 0.356 & 0.356 & 3\\ 
    Bibit     & 0.006 & 0.007 & 0.006 & 0.018 & 36315\\ 
    Bibit2    & 0.007 & 0.008 & 0.007 & 0.022 & 36315\\ 
    Bimax     & 0.000 & 0.005 & 0.005 & 0.006 & 100\\
    iBBiG(3)  & 0.198 & 0.205 & 0.198 & 0.296 & 3\\
    iBBiG(10) & 0.098 & 0.289 & 0.293 & 0.117 & 10\\
    GBC(3)    & 0.182 & 0.264 & 0.231 & 0.272 & 3\\
    GBC(10)   & 0.073 & 0.249 & 0.226 & 0.093 & 10\\
    \hline
    \end{tabular} 
    \label{tab:snp} 
 \end{table}

 Table \ref{tab:snp} shows that the results from our analysis. We can see that BiSSLB had the highest CS, CE, recovery, and relevance among all the binary biclustering methods. Moreover, BiSSLB accurately recovered the three biclusters ($\widehat{K}=3$) with the initial overestimate of $K^{\star} = 10$. On the other hand, iBBiG(10) and GBC(10) did \textit{not} shrink the number of estimated biclusters. Therefore, we also fit iBBiG and GBC with an initial guess of three biclusters, denoted by iBBiG(3) and GBC(3). This increased the performance metrics for iBBiG and GBC, but they still had lower CS, CE, recovery, and relevance than BiSSLB. Bibit, Bibit2, and Bimax all grossly overestimated the number of biclusters, leading to poor performance scores. On this specific dataset, BiSSLB was the only method to automatically learn the correct number of biclusters, \textit{and} it achieved the highest performance metrics even when competing methods were initialized with the true number of biclusters.  

\begin{figure}[t!]
    \centering
    \includegraphics[width=.9\linewidth]{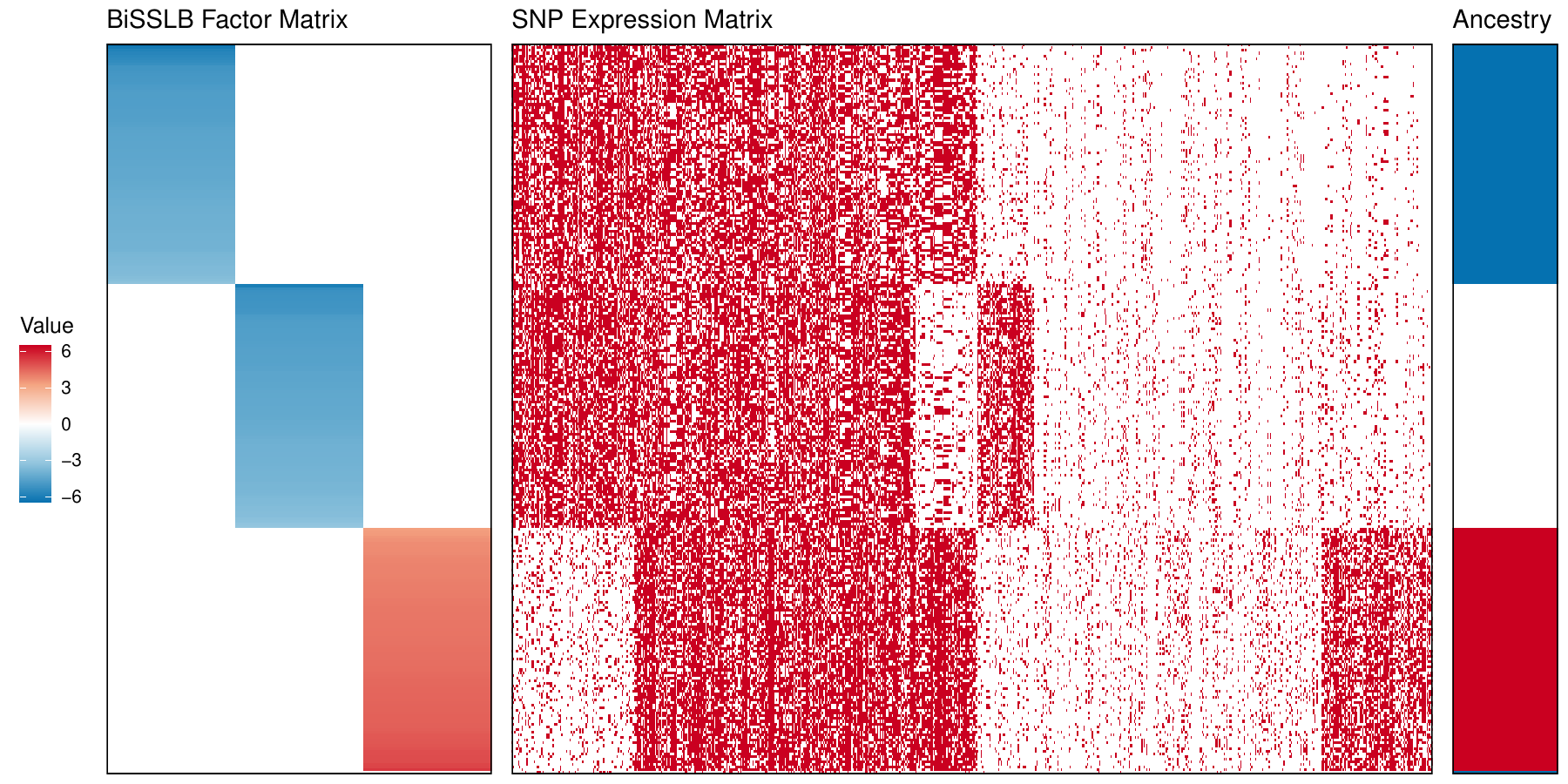}
    \caption{Results from fitting BiSSLB to the HapMap data. \textbf{Left panel}: the reordered latent factor matrix $\mathbf{A}$ with the three biclusters that BiSSLB found. Each row represents one sample. \textbf{Middle panel}: the reordered SNP data matrix where the rows correspond to the samples and the columns correspond to the SNP genotypes. A red square indicates a mutant (``1''), and a white square indicates a wild-type and (``0''). \textbf{Right panel}: the ancestry information (Blue = Caucasians, White = Asians, Red = Africans).}
    \label{fig:snp_bisslb}
\end{figure}

Figure \ref{fig:snp_bisslb} plots the reordered estimated latent matrix $\mathbf{A}$ (left panel) and the corresponding reordered SNP genotype matrix (middle panel). The SNP data appears to be very noisy, with many isolated ``1''s (denoted by red squares), as well as noisy clusters containing many ``0''s (denoted by white squares). This noisiness may explain why Table \ref{tab:snp} shows relatively lower performance scores  and why Bibit, Bibit2, Bimax, iBBiG, and GBC all struggled to recover the true bicluster count. Despite the high level of noise in this dataset, Figure \ref{fig:snp_bisslb} indicates that BiSSLB was able to recover the true biclusters corresponding to the different ethnicities (right panel of Figure \ref{fig:snp_bisslb}). The estimated latent matrix $\mathbf{A}$ (left panel) also shows that the submatrices associated with Caucasian and Asian individuals had more similar values (colored in blue), than the submatrix associated with African individuals (colored in red). This indicates that Caucasians and Asians share more similar SNP regions than with Africans, a finding that is consistent with the middle panel of Figure \ref{fig:snp_bisslb} and with previous analyses of this dataset \citep{lee2014biclustering}. 

\subsection{Homo Sapiens protein-protein interaction analysis}\label{ssec:ppi}

We also applied our method to PPI analysis, following the framework of \citet{sriwastava2023rubic}. Specifically, we used data on human (\textit{Homo sapiens}) PPIs from the Database of Interacting Proteins (DIP), which comprises of 6247 records of interactions \citep{salwinski2004database}. This dataset includes RefSeq identifiers, UniProt Knowledgebase (UniProtKB) accession numbers, and DIP identifiers, each serving different roles in annotating protein sequences and interactions \citep{salwinski2004database}. To remove redundant entries, we relied on UniProtKB accession numbers \citep{UniProt2016}, as they are stable across database updates and merges and are well-suited for subsequent mapping to gene information.

After preprocessing, the dataset contained 3851 unique interaction entries. We represented these data as a $3851 \times 3851$ binary matrix, where a ``1'' indicates an interaction between two proteins and ``0'' denotes no known interaction. We then fit the different biclustering methods to the data. For BiSSLB, Bimax, and iBBiG, we performed multiple runs to identify an appropriate guess of the number of biclusters. Specifically, we increased the initial guess $K^{\star}$ until the results were relatively stable and further increases in $K^{\star}$ did not result in a different number of estimated biclusters. We were not able to fit GBC beyond an initial guess of $K^{\star}=50$ due to its computational intensity. This may be because GBC uses an EM algorithm which we found to be much slower than the other methods. Hence, we set $K^{\star} = 50$ for GBC.

\begin{table}[t!]
    \centering
    \caption{Comparison of different biclustering methods on the Homo Sapians PPI dataset. The last column $\widehat{K}$ is the estimated number of biclusters for each method.}
    \begin{tabular}{lccccc}
    \hline
    \textbf{Method} & \textbf{AUC} & \textbf{AUPR} & $\widehat{\textbf{K}}$ \\ 
    \hline
    BiSSLB  & 0.965 & 0.684 & 311 \\
    Bibit   & 0.756 & 0.514 & 2046 \\
    Bibit2  & 0.821 & 0.092 & 2046 \\
    Bimax   & 0.756 & 0.514 & 2729 \\
    iBBiG   & 0.775 & 0.492 &  1000 \\
    GBC     & 0.878 & 0.089 & 50 \\
    \hline
    \end{tabular}
    \label{tab:ppi}
\end{table}

In this PPI dataset, we do not know the ground truth biclusters. Therefore, we compared the different methods based on predictive accuracy.  
Table \ref{tab:ppi} presents the area under the Receiver Operating Characteristic curve (AUC), the area under the precision-recall curve (AUPR), and the estimated number of biclusters $\widehat{K}$ for the different binary biclustering algorithms. Table \ref{tab:ppi} shows that BiSSLB achieved the best overall performance, with the highest AUC (0.965) and AUPR (0.684). Additionally, BiSSLB estimated a much smaller number of biclusters ($\widehat{K} = 311$) than Bibit, Bibit2, Bimax, or iBBiG, all of which estimated bicluster counts in the thousands. Thus, BiSSLB learned a much more compact representation without sacrificing any predictive accuracy. GBC initialized with 50 biclusters had the second highest AUC, but its AUPR was also the lowest. However, we stress that we were unable to fit GBC with more than 50 biclusters. 

\begin{figure}[t!]
    \centering
    \includegraphics[width=.8\linewidth]{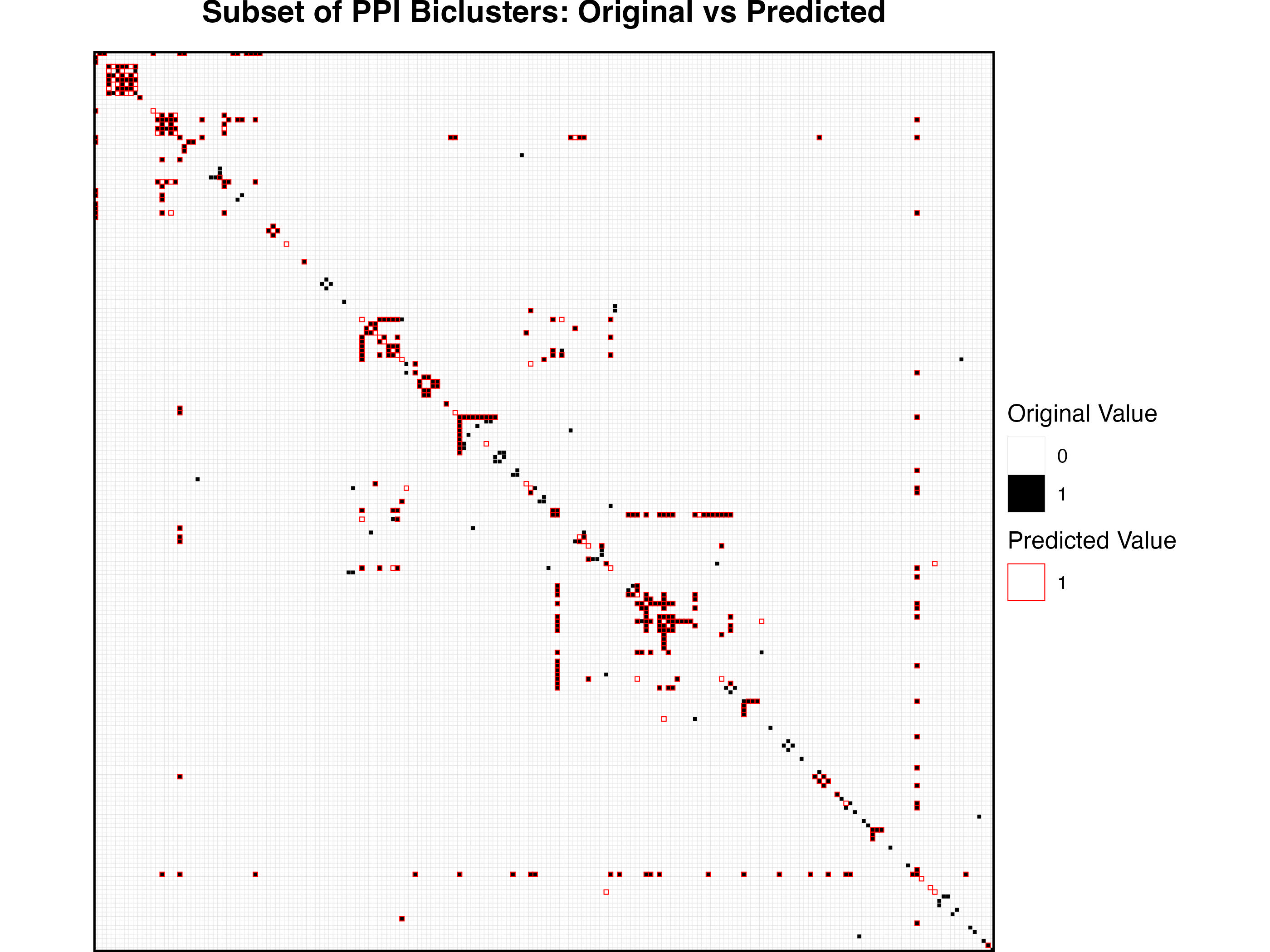}
    \caption{A $200 \times 200$ submatrix of the original PPI data, where black cells indicate observed interactions (``1'') and white cells represent non-interactions (``0''). The biclusters estimated by BiSSLB are overlaid in red.}
    \label{fig:ppi_bisslb}
\end{figure}

Figure \ref{fig:ppi_bisslb} plots a representative $200 \times 200$ submatrix of the original PPI data, with the estimated BiSSLB biclusters (in red) overlaid on the observed data submatrix (black squares for known interactions and white squares for non-interactions). We can see that the patterns unveiled by BiSSLB aligned well with the observed data. Moreover, Figure \ref{fig:ppi_bisslb} shows that the biclusters found by BiSSLB were of varying sizes and naturally accommodated noisy structures in the data. Overall, our analysis suggests that BiSSLB exhibited the greatest robustness to noise on this PPI dataset.

\section{Conclusion}\label{sec:conclusion}

We proposed BiSSLB, a new Bayesian biclustering algorithm that extends the work of \citet{moran2021spike} to binary data matrices. By leveraging spike-and-slab priors within a Bayesian nonparametric framework, BiSSLB effectively captures latent clustering structures and adaptively learns the number of biclusters from the data. This automatic adaptation mitigates the overfitting/underfitting issues that commonly arise in other binary biclustering methods. We further introduced an efficient coordinate ascent algorithm for BiSSLB which overcomes several drawbacks with the EM algorithm (e.g., slow convergence and local entrapment). Through simulation studies and real analyses of SNP and PPI data, we demonstrated the superior performance of BiSSLB over several other state-of-the-art methods for biclustering noisy binary datasets.

A natural extension of BiSSLB would be to biclustering for other types of discrete data. For example, in cancer genomics, mutational signatures corresponding to distinct processes (e.g., damaged DNA repair mechanisms and environmental mutagens) are defined by the count frequencies of specific mutation types \citep{Alexandrov2013Nature}. BiSSLB can be extended to mutational signatures analysis by replacing the Bernoulli likelihood with a Poisson or negative binomial model and connecting our latent matrix factorization model to the expected counts through a log link function. It is also of interest to perform biclustering on mixed-type data, i.e., datasets consisting of both continuous and discrete variables \citep{li2020bayesian}. BiSSLB can also be extended in this direction by decomposing the likelihood function into different parts associated with different data types (e.g., count, binary, and continuous data) and then linking our latent matrix factorization model to the expected values of the original data through appropriate link functions.

\appendix

\numberwithin{equation}{section} 

\renewcommand{\thefigure}{A.\arabic{figure}}
\setcounter{figure}{0} 

\section{Details and Derivations for the BiSSLB algorithm} \label{Sec:App-Algorithm}

\subsection{Proximal gradient updates for $\mathbf{A}$ and $\mathbf{B}$} 

As discussed in the main article, We use proximal gradient steps to update $\mathbf{A}$ and $\mathbf{B}$ directly in our BiSSLB algorithm. This allows us to bypass the EM algorithm. In what follows, we describe the update for $\mathbf{A}$ in detail (holding $\mathbf{B}$ fixed at its current value), noting that the update for $\mathbf{B}$ (holding $\mathbf{A}$ fixed) is very similar. Recall that $K^{\star}$ is the pseudo-upper bound on the number of biclusters $K$.

 We first define the spike-and-slab lasso (SSL) penalty function for the $k$th column $\mathbf{a}_k$ of $\mathbf{A}$ in (2), where the $i$th entry of $\mathbf{a}_k$ is denoted by $a_{ik}$. We denote this penalty function as $\operatorname{pen}(\mathbf{a}_k)$. By marginalizing over the binary indicator variables $\widetilde{\boldsymbol{\Gamma}}$ and the Bernoulli probabilities $\widetilde{\boldsymbol{\theta}}$ in (4)-(5), we arrive at the following marginal prior for $\mathbf{a}_k$:
\begin{equation*} 
    \pi(\boldsymbol a_{k})=\int \prod_{i=1}^{I} \left[ (1-\widetilde{\theta}_{(k)}) \psi \left(a_{ik} \mid \widetilde{\lambda}_0 \right)+\widetilde{\theta}_{(k)}\psi \left(a_{ik} \mid \widetilde{\lambda}_1 \right)\right] \pi(\widetilde{\theta}_{(k)}) d \widetilde{\theta}_{(k)}, 
\end{equation*}
where $\pi(\widetilde{\theta}_{(k)})$ denotes the density function for the hyperprior on $\widetilde{\theta}_{(k)}$.  This was described by \citet{rovckova2018spike} as the non-separable spike-and-slab lasso (NSSL) penalty. 

Since we aim to find a MAP estimator for the BiSSLB posterior, the log prior distribution $\log \pi(\mathbf{a}_k)$ can be thought of as a penalty function on $\mathbf{a}_k$. However, similar to \citet{rovckova2018spike}, we first modify this penalty to be centered at zero by adding a constant to ensure that $\operatorname{pen}( \mathbf{0}_{K^{\star}}) = 0$, i.e., 
\begin{equation} \label{BiSSLB-penak}
    \operatorname{pen}(\boldsymbol a_k) = \log \frac{\pi\left(\boldsymbol{a}_k\right)}{\pi\left(\boldsymbol{0}_{K^\star}\right)} = \sum_{i=1}^I-\widetilde{\lambda}_1 \left|a_{i k}\right|+\log \left[ \frac{p^*(0 ; \widetilde{\theta}_{ik}, \widetilde{\lambda}_0, \widetilde{\lambda}_1)}{p^* (a_{i k} ; \widetilde{\theta}_{ik}, \widetilde{\lambda}_0, \widetilde{\lambda}_1 )} \right], 
\end{equation}
where $\mathbf{0}_{K^{\star}}$ is the $K^{\star}$-dimensional zero vector, and the function $p^{\star}(\cdot ~;~ \cdot, \cdot, \cdot)$ is defined as
\begin{equation} \label{BiSSLB-p-star}
    p^*(x; \theta, \xi_0, \xi_1) =\theta \psi \left(x\mid \xi_1 \right) /\left[(1-\theta) \psi \left(x \mid \lambda_0 \right) + \theta \psi \left(x \mid \lambda_1 \right) \right],
\end{equation}
and 
\begin{equation} \label{BiSSLB-thetaik}
    \widetilde{\theta}_{ik}= \mathbb{E} [\widetilde{\theta}_{(k)} \mid \boldsymbol{a}_{k \backslash i} ], 
\end{equation}
where $\boldsymbol{a}_{k \backslash i}$ is defined as $\boldsymbol{a}_k$ with the $i$th element removed. Taking the derivative of $\text{pen}(\mathbf{a}_k)$ with respect to $| a_{ik}|$ yields
\begin{equation*}
    \frac{\partial \operatorname{pen}(\boldsymbol a_k)}{\partial | a_{ik}|} = -\lambda^{\star} (a_{ik}; \widetilde{\theta}_{ik}, \widetilde{\lambda}_0, \widetilde{\lambda}_1 ),
\end{equation*}
where the function $\lambda^{\star}(\cdot ~;~ \cdot, \cdot, \cdot)$ is defined as
\begin{equation} \label{BiSSLB-lambdastar}
\lambda^{\star} (x ; \theta, \xi_0, \xi_1 ) = \xi_1 p^{\star} ( x ; \theta, \xi_0, \xi_1 ) + \xi_0 \left[1-p^* (x ; \theta, \xi_0, \xi_1) \right].
\end{equation}
This allows us to represent the penalty term \eqref{BiSSLB-penak} more succinctly as 
\begin{equation}\label{BiSSLB-penalty-ak}
    \operatorname{pen}(\boldsymbol a_k) = \sum_{i = 1}^{I}-\lambda^{\star}(a_{ik}; \widetilde{\theta}_{ik}, \widetilde{\lambda}_0, \widetilde{\lambda}_1 )|a_{ik}|. 
\end{equation}
Although \eqref{BiSSLB-penalty-ak} looks similar to the lasso penalty function of \citet{tibshirani1996regression}, there are some crucial differences. Whereas the lasso uses a single penalty parameter $\lambda$ for all coordinates, \eqref{BiSSLB-penalty-ak} applies a \emph{different} penalty term $\lambda^{\star} (a_{ik}; \widetilde{\theta}_{ik}, \widetilde{\lambda}_0, \widetilde{\lambda}_1)$ for each $a_{ik}$. In particular, if $|a_{ik}|$ is small, then $\lambda^{\star}(a_{ik}; \widetilde{\theta}_{ik} \widetilde{\lambda}_0, \widetilde{\lambda}_1)$ will be large, and if $|a_{ik}|$ is large, then $\lambda^{\star}(a_{ik}; \widetilde{\theta}_{ik}, \widetilde{\lambda}_0, \widetilde{\lambda}_1)$ will be small  \citep{rovckova2018spike,moran2019variance,moran2021spike}. This adaptive shrinkage enables us to more easily find biclusters, since the large magnitude entries of $\mathbf{A}$ and $\mathbf{B}$ are not penalized (or shrunk) too much.

Suppose that $\mathbf{B}$ is fixed at its current value. Based on \eqref{BiSSLB-penalty-ak}, maximizing the BiSSLB log-posterior (7) with respect to $\mathbf{A}$ is equivalent  to miniizing the following objective function:
\begin{equation}\label{eq:BiSSLB-proximal}
    \widehat{\mathbf{A}} = \underset{\mathbf{A}\in \mathbb{R}^{I\times K^{\star}}}{\arg\min} \ f(\mathbf{A})+h(\mathbf{A}), 
\end{equation}
where $f(\mathbf{A})$ denotes the differentiable negative log-likelihood function, i.e.,
\begin{equation} \label{eq:BiSSLB-proximal-fA}
    f(\mathbf{A}) = \sum_{i=1}^{I} \sum_{j=1}^{J} \left[\log\left(1+\exp\left(\boldsymbol{\mu}\mathbf{1}^{\top} + \mathbf{AB^{\top}}\right)\right) - \mathbf{Y} \odot\left(\mathbf{AB^{\top}}\right)\right]_{i j}, 
\end{equation}
while $h(\mathbf{A})$ denotes the \emph{non}-differentiable function corresponding to the negative SSL penalty, i.e.,
\begin{equation} \label{BiSSLB-proximal-hA}
    h(\mathbf{A}) = \sum_{k=1}^{K^{\star}} - \operatorname{pen}(\boldsymbol a_k) = \sum_{k=1}^{K^{\star}}\sum_{i=1}^{I} \lambda^{\star}(a_{ik}; \widetilde{\theta}_{ik}, \widetilde{\lambda}_0, \widetilde{\lambda}_1) |a_{ik}|. 
\end{equation}
The decomposition of the objective function in \eqref{eq:BiSSLB-proximal} into differentiable and non-differentiable parts suggests a proximal gradient descent update for $\mathbf{A}$. We define the proximal operator for $\mathbf{U} = (u_{ik}) \in \mathbb{R}^{I \times K^{\star}}$ as the solution to the following problem: 
\begin{equation} \label{eq:BiSSLB-prox} 
    \operatorname{prox}_\eta(\mathbf{U}) =  \underset{\mathbf{X}}{\arg\min} \ \frac{1}{2 \eta}\|\mathbf{U} - \mathbf{X} \|_\text{F}^2 + h(\mathbf{X}),
\end{equation}
where $\eta$ is a step size (or learning rate) and $h(\cdot)$ is the function in \eqref{BiSSLB-proximal-hA}. 

Then the proximal gradient update for $\mathbf{A}$ at the $t$th iteration, $t \geq 1$, is
\begin{equation} \label{BiSSLB-proximal-operator}
    \mathbf{A}^{(t)} =  \operatorname{prox}_\eta\left(\mathbf{A}^{(t-1)} - \eta \nabla f(\mathbf{A}^{(t-1)})\right),
\end{equation}
where $\nabla f(\mathbf{A})$ is the gradient of \eqref{eq:BiSSLB-proximal-fA} with respect to $\mathbf{A}$, i.e., 
\begin{equation} \label{BiSSLB-gradient-fA}
    \nabla f(\mathbf{A}) = \left(\mathbf{W} - \mathbf{Y}\right) \mathbf{B}, 
\end{equation}
where $\mathbf{W} = (w_{ij})$ is a matrix whose $(i,j)$th entry is $w_{ij} = 1 / \{ 1 + \exp (- \mu_i - \mathbf{a}_i^\top \mathbf{b}_j ) \}$, and $\mathbf{a}_i$ and $\mathbf{b}_j$ are respectively the $i$th row of $\mathbf{A}$ and the $j$th row of $\mathbf{B}$. 

The step size parameter $\eta$ comes from the Taylor expansion of the objective function, which plays a similar role as the variance parameter ($\sigma^2$) in Gaussian error models for continuous data, as considered in \citet{moran2021spike}. In contrast to SSLB, which updates $\sigma^2$ iteratively via the EM algorithm, we adopt a fixed step size in BiSSLB. Specifically, $\eta$ is selected through a grid search over values such as $\{ 10^{-2}, 10^{-4}, 10^{-6}, \ldots \}$. Consequently, $\eta$ serves as a tuning hyperparameter that controls the aggressiveness of the optimization procedure.  

Since each term $\lambda^{\star}(a_{ik}; \widetilde{\theta}_{ik}, \widetilde{\lambda}_0, \widetilde{\lambda}_1)$ in \eqref{BiSSLB-proximal-hA} depends on $a_{ik}$, the objective function \eqref{eq:BiSSLB-prox} is highly nonconvex, and thus, it may be difficult to find the global mode of the proximal operator in \eqref{BiSSLB-proximal-operator}. However,
\citet{rovckova2018spike} gave a refined characterization of the global mode under the SSL prior for high-dimensional linear regression, which we adapt for our algorithm. This refined characterization of the global mode allows us to eliminate many suboptimal local modes in each iteration, thereby greatly improving the chances that our algorithm converges to a local mode that is close to or equal to the global mode.

We have the following proposition based on Theorems 3 and 4 of \citet{rovckova2018spike}.

\begin{proposition}\label{BiSSLB-proposition1}
    \textit{Let $\widehat{\mathbf{A}}$ denote the global mode of the proximal operator \eqref{BiSSLB-proximal-operator}, with $(i,k)$th entry $\widehat{a}_{ik}$, and let $\widetilde{\theta}_{ik}$ be defined as in \eqref{BiSSLB-thetaik}. Then}
    $$
    \widehat{a}_{ik}= 
    \begin{cases} 
        0, & \text { when }\left|z_{ik}\right| \leq \Delta, \\ 
        \left[\left|z_{ik}\right|-\eta \lambda^{\star}(\widehat{a}_{ik}; \widetilde{\theta}_{ik}, \widetilde{\lambda}_0, \widetilde{\lambda}_1 )\right]_{+} \operatorname{sign}\left(z_{ik}\right), & \text { when }\left|z_{ik}\right|>\Delta,
    \end{cases}
    $$
    \textit{where $\mathbf{Z} = \mathbf{A}^{(t-1)} - \eta \nabla f(\mathbf{A}^{(t-1)})$ with $(i,k)$th entry $z_{ik}$, $
    \Delta \equiv \inf _{t>0} [t / 2-\eta \operatorname{pen} ( t \mid \widetilde{\theta}_{ik} ) / t ]$, and $x_{+} = \max \{ 0, x \}$.}
    
    \textit{Moreover, define the function $g(\cdot ~;~ \cdot, \cdot, \cdot)$ as 
    \begin{equation}  \label{BiSSLB-g-function}
    g(x ; \theta, \xi_0, \xi_1) = [ \lambda^{\star}(x; \theta, \xi_0, \xi_1) - \lambda_1 ]^2 + (2 / \eta) \log [p^*(x; \theta, \xi_0, \xi_1)].
    \end{equation}
    When $\sqrt{\eta}\left(\lambda_0-\lambda_1\right)>2$ and $g(0 ; \widetilde{\theta}_{ik}, \widetilde{\lambda}_0, \widetilde{\lambda}_1 )>0$, the threshold $\Delta$ is bounded by}
    $$
    \Delta^L<\Delta<\Delta^U, 
    $$
    \textit{where}
    $$
    \begin{aligned}
        & \Delta^L=\sqrt{2 \eta \log \left[1 / p^*(0 ; \widetilde{\theta}_{ik}, \widetilde{\lambda}_0, \widetilde{\lambda}_1 )\right]-\eta^2 d}+\eta \widetilde{\lambda}_1,  \\
        & \Delta^U=\sqrt{2 \eta \log \left[1 / p^*(0 ; \widetilde{\theta}_{ik}, \widetilde{\lambda}_0, \widetilde{\lambda}_1 )\right]}+\eta \widetilde{\lambda}_1,
    \end{aligned}
    $$
    \textit{and}
    $$
    0<d<\frac{2}{\eta}-\left(\frac{1}{\eta\left(\widetilde{\lambda}_0-\widetilde{\lambda}_1\right)}-\sqrt{\frac{2}{\eta}}\right)^2.
    $$
\end{proposition}
\citet{rovckova2018spike} noted that when the spike hyperparameter $\widetilde{\lambda}_0$ is large, $\Delta^{L} \approx \Delta^{U}$. Further, as we explain in Section \ref{sparsity-updates}, $\widetilde{\theta}_{ik}$ can be approximated by $\widetilde{\tau}_k$, where $\widetilde{\tau}_k = \mathbb{E}[ \widetilde{\theta}_{(k)} \mid \mathbf{a}_k ]$. Thus, we can incorporate the refined characterization of the global mode in Proposition \ref{BiSSLB-proposition1} into our proximal update \eqref{BiSSLB-proximal-operator} as follows. 

Define the generalized soft-thresholding operator \citep{mazumder2011sparsenet} as
\begin{equation} \label{eq:BiSSLB-GST}
\widetilde{S}(z, \lambda, \Delta)=\frac{1}{n}(|z|-\lambda)_{+} \operatorname{sign}(z) \mathbb{I}(|z|>\Delta) .
\end{equation}
Then the proximal update \eqref{BiSSLB-proximal-operator} for $\mathbf{A}$ at the $t$th iteration can be succinctly written as
\begin{equation} \label{BiSSLB-refined-A-update}
a^{(t)}_{ik} \gets \widetilde{S}\left(z_{ik}, ~ \eta \lambda^{\star} (a^{(t-1)}_{ik} ; \widetilde{\tau}_k^{(t-1)} ; \widetilde{\lambda}_0, \widetilde{\lambda}_1 ), ~ \Delta^{U}\right), \quad i = 1, \ldots, I, ~~ k = 1, \ldots, K^{\star},
\end{equation}
where 
\begin{equation*} 
\Delta^U = 
\begin{cases}
    \sqrt{2 \eta \log \left[1 / p^* (0 ; \widetilde{\tau}_k^{(t-1)}, \widetilde{\lambda}_0, \widetilde{\lambda}_1 )\right]}+\eta \lambda_1, & \text { if } g(0 ; \widetilde{\tau}_k^{(t-1)}, \widetilde{\lambda}_0, \widetilde{\lambda}_1)>0, \\ 
    \eta \lambda^{\star}(0 ; \widetilde{\tau}_k^{(t-1)}, \widetilde{\lambda}_0, \widetilde{\lambda}_1), & \text { otherwise},
\end{cases}
\end{equation*}
and $p^{\star}(\cdot ~;~ \cdot, \cdot, \cdot)$ is defined as in \eqref{BiSSLB-p-star}, $\lambda^{\star}(\cdot ~;~ \cdot, \cdot, \cdot)$ is defined as in \eqref{BiSSLB-lambdastar}, and $g(\cdot ~;~ \cdot, \cdot, \cdot)$ is defined as in \eqref{BiSSLB-g-function}. Note that in \eqref{BiSSLB-refined-A-update}, we have written $\lambda^{\star}(a^{(t-1)}_{ik}; \widetilde{\tau}_k^{(t-1)}, \widetilde{\lambda}_0, \widetilde{\lambda}_1)$ to explicitly indicate that this refined update is a function dependent on both $a^{(t-1)}_{ik}$ and $\widetilde{\tau}_{k}^{(t-1)}$ from the $(t-1)$st iteration of the algorithm. 

It is worth stressing that the update \eqref{BiSSLB-refined-A-update} for $\mathbf{A}$ is a combination of both soft-thresholding and hard-thresholding. In particular, many suboptimal local modes for $\mathbf{A}$ are eliminated through the threshold $\Delta^U$. To further accelerate convergence in practice, we apply the momentum-based update in the fast iterative shrinkage-thresholding (FISTA) algorithm \citep{beck2009fista}. Namely, after initializing $\mathbf{A}^{(0)} = \mathbf{A}^{(1)}$, we update $\mathbf{A}^{(t)}$ in each $t$th iteration, $t \geq 2$, as
\begin{equation} \label{BiSSLB-A-update}
\boxed{
    \begin{aligned}
        & \mathbf{A}_m \leftarrow  \mathbf{A}^{(t-1)} + \frac{t-2}{t+1}\left(\mathbf{A}^{(t-1)} - \mathbf{A}^{(t-2)}\right), \\
        & \mathbf{Z} \leftarrow \mathbf{A}_m - \eta \nabla f(\mathbf{A}_m), \\ 
        & \text{Update } \mathbf{A}^{(t)} \text{ as in } \eqref{BiSSLB-refined-A-update},
    \end{aligned}}
\end{equation}
where $\nabla f(\cdot)$ is the gradient in \eqref{BiSSLB-gradient-fA}. 

The updates for $\mathbf{B}$ are analogous to those for $\mathbf{A}$. Suppose that $\mathbf{A}$ is fixed at its current value. Let $\tau_k = \mathbb{E}[ \theta_{(k)} \mid \mathbf{b}_k ]$. The proximal update for $\mathbf{B}$ at the $t$th iteration is
\begin{equation} \label{BiSSLB-refined-B-update}
b^{(t)}_{jk} \gets \widetilde{S}\left(z_{jk}, \eta \lambda^{\star} (b^{(t-1)}_{jk} ; \tau_k^{(t-1)}, \lambda_0, \lambda_1 ), \Gamma^{U}\right),  \qquad j = 1, \ldots, J, ~~ k = 1, \ldots, K^{\star},
\end{equation}
where $\widetilde{S}(\cdot, \cdot, \cdot)$ is the generalized soft-thresholding operator \eqref{eq:BiSSLB-GST}, $z_{jk}$ is the $(j,k)$th entry of $\mathbf{Z}$ in the proximal update $\mathbf{B}^{(t)} = \operatorname{prox}_{\eta}(\mathbf{Z})$, and
\begin{equation*}
\Gamma^U = 
\begin{cases}
    \sqrt{2 \eta \log \left[1 / p^*(0 ;  \tau_k^{(t-1)}, \lambda_0, \lambda_1)\right]}+\eta \lambda_1, & \text { if } g(0 ; \tau_{k}^{(t-1)}, \lambda_0, \lambda_1)>0, \\ 
    \eta \lambda^{\star}(0 ;  \tau_{k}^{(t-1)}, \lambda_0, \lambda_1), & \text { otherwise},
\end{cases}
\end{equation*}
 $p^{\star}(\cdot ~; ~ \cdot, \cdot, \cdot)$ is defined in in \eqref{BiSSLB-p-star}, $\lambda^{\star}(\cdot ~;~ \cdot, \cdot, \cdot)$ is defined as in \eqref{BiSSLB-lambdastar}, and $g(\cdot ~;~ \cdot, \cdot, \cdot)$ is defined as in \eqref{BiSSLB-g-function}.

Similarly as with the update of $\mathbf{A}$, the refined update \eqref{BiSSLB-refined-B-update} for $\mathbf{B}$ is a combination of soft-thresholding and hard-thresholding, with suboptimal modes eliminated by the threshold $\Gamma^{U}$. After initializing $\mathbf{B}^{(0)}=\mathbf{B}^{(1)}$, we update $\mathbf{B}^{(t)}$ in each $t$th iteration, $t \geq 2$, as
\begin{equation} \label{BiSSLB-B-update}
\boxed{
    \begin{aligned}
        & \mathbf{B}_m \leftarrow  \mathbf{B}^{(t-1)} + \frac{t-2}{t+1}\left(\mathbf{B}^{(t-1)} - \mathbf{B}^{(t-2)}\right), \\
        & \mathbf{Z} \leftarrow \mathbf{B}_m - \eta \nabla f(\mathbf{B}_m), \\ 
        & \text{Update } \mathbf{B}^{(t)} \text{ as in } \eqref{BiSSLB-refined-B-update},
    \end{aligned}}
\end{equation}
where $\nabla f(\mathbf{B}) = (\mathbf{W} -\mathbf{Y}) \mathbf{A}$, and $\mathbf{W}$ is the same matrix as in \eqref{BiSSLB-gradient-fA}.


\subsection{Update for the location parameter} 
Under the flat prior $\pi(\boldsymbol{\mu}) \propto 1$, we aim to find  $\boldsymbol{\mu}$ which maximizes $g(\boldsymbol{\mu})$, where
\begin{equation} \label{J-mu}
    g\left( \boldsymbol{\mu} \right) = 
        \sum_{i=1}^{I} \sum_{j=1}^{J} \left[\mathbf{Y} \odot \boldsymbol{\mu}\mathbf{1}^{\top} - \log\left(1+\exp\left(\boldsymbol{\mu}\mathbf{1}^{\top} + \mathbf{AB^{\top}}\right)\right)\right]_{i j}.
\end{equation}
Let $\mathbf{A}$ and $\mathbf{B}$ be fixed at their current values. To derive the update for $\boldsymbol{\mu}$ in each $t$th iteration of our coordinate algorithm, we employ a coordinatewise minorize-maximization (MM) algorithm. Let $\mu_i$ be the $i$th entry in $\boldsymbol{\mu}$. Maximizing \eqref{J-mu} with respect to $\boldsymbol{\mu}$ is equivalent to maximizing the following objective with respect to $\mu_i$ for each $i = 1, \ldots, I$:
\begin{equation} \label{ell-mui}
    \ell(\mu_i)=\sum_{j=1}^J y_{ij} \mu_i-\log \left(1+e^{\mu_i+\boldsymbol{a}_i^{\top} \boldsymbol{b}_j} \right).
\end{equation}
The gradient and the Hessian of $\ell(\mu_i)$ respectively are
\begin{equation*}
    \nabla \ell(\mu_i) = \sum_{j=1}^{J} y_{ij} - p_{ij} \quad \text{and} \quad  \nabla^2 \ell(\mu_i)=-\sum_{j=1}^J p_{ij}\left(1-p_{ij}\right),
\end{equation*}
where $p_{ij} = 1/ \{ 1+\exp(-(\mu_i + \mathbf{a}_i^\top \mathbf{b}_j) \}$. Since \eqref{ell-mui} is concave and twice-differentiable and $\nabla^2 \ell(\mu_i)$ can be lower-bounded by $-J/4$, we can lower-bound $\ell(\mu_i)$ as 
\begin{equation*}
    \ell(\mu_i) \geq \ell\left(\mu_i^{(t-1)}\right)+\nabla \ell\left(\mu_i^{(t-1)}\right)\left(\mu_i-\mu_i^{(t-1)}\right) - \frac{J}{8}\left(\mu_i-\mu_i^{(t-1)}\right)^{2}.
\end{equation*}
Thus, we can define the minorizing function $g(\mu_i \mid \mu_i^{(t)})$ for $\ell(\mu_i)$ as
\begin{equation} \label{minorizing}
    g\left(\mu_i \mid \mu_i^{(t-1)}\right)=\ell\left(\mu_i^{(t-1)}\right) + \left( \sum_{j=1}^{J} y_{ij} - p_{ij}^{(t-1)} \right)  \left(\mu_i-\mu_i^{(t-1)}\right) - \frac{J}{8}\left(\mu_i-\mu_i^{(t-1)}\right)^{2}. 
\end{equation}
Maximizing \eqref{minorizing} with respect to $\mu_i$ leads to the following update for each $\mu_i, i = 1, \ldots, I$, in the $t$th iteration:
\begin{equation}
        \mu_i^{(t)}  \leftarrow \mu_i^{(t-1)}+\frac{4}{J}\left(\sum_{j=1}^J y_{ij}-p_{ij}^{(t-1)}\right).
\end{equation}

\subsection[Updates for tau and tau]{Updates for $\widetilde{\tau}$ and $\tau$} \label{sparsity-updates} 
Based on Proposition \ref{BiSSLB-proposition1}, the \emph{exact} updates for $a_{ik}$ and $b_{jk}$ would have required us to evaluate
\begin{equation} \label{BiSSLB-exact-theta-expectations}
    \widetilde{\theta}_{ik} = \mathbb{E}[\widetilde{\theta}_{(k)} \mid \boldsymbol{a}_{k \setminus i}] \quad \text{and} \quad \theta_{jk} = \mathbb{E}[{\theta}_{(k)} \mid \boldsymbol{b}_{k \setminus j}],
\end{equation}
where $\boldsymbol{a}_{k \setminus i}$ is the $k$th column of $\mathbf{A}$ with the $i$th element removed and while $\boldsymbol{b}_{k \setminus j}$ is the $k$th column of $\mathbf{B}$ with the $j$th element removed. However, when $I$ and $J$ are large, \cite{rovckova2018spike} observed that the conditional expectations \eqref{BiSSLB-exact-theta-expectations} are very close to $\mathbb{E}[\widetilde{\theta}_{(k)} \mid \boldsymbol{a}_{k}]$ and $\mathbb{E}[{\theta}_{(k)} \mid \boldsymbol{b}_{k}]$, respectively. Therefore, for simplicity and practical implementation, we replace the expectations in \eqref{BiSSLB-exact-theta-expectations} with their approximations,
\begin{equation} \label{BiSSLB-approximate-theta-expectations}
  \widetilde{\tau}_k = \mathbb{E}[\widetilde{\theta}_{(k)} \mid \boldsymbol{a}_{k}] \quad \text{and} \quad \tau_k = \mathbb{E}[{\theta}_{(k)} \mid \boldsymbol{b}_{k}].  
\end{equation}
in the refined updates \eqref{BiSSLB-refined-A-update} and \eqref{BiSSLB-refined-B-update} for $a_{ik}$ and $b_{jk}$.
These conditional expectations do not have closed-form expressions. However, we can approximate them as follows. First, note that the IBP priors with intensity parameters $\widetilde{\alpha} > 0$ and $\alpha > 0$ arise from the beta-Bernoulli priors,
\begin{equation*}
    \begin{aligned}
    & \pi(\widetilde{\gamma}_{jk} \mid \widetilde{\theta}_k) \overset{\text{ind}}{\sim} \text{Bernoulli}(\widetilde{\theta}_k), ~~~ \pi(\widetilde{\theta}_k) \sim \text{Beta}\left( \frac{\widetilde{\alpha}}{K}, 1 \right),~~~ k = 1, 2, \ldots  \\
    & \pi(\gamma_{jk} \mid \theta_k ) \overset{\text{ind}}{\sim} \text{Bernoulli}(\theta_k), ~~~ \pi(\theta_k) \sim \text{Beta}\left( \frac{\alpha}{K}, 1 \right), ~~~ k= 1, 2, \ldots
    \end{aligned}
\end{equation*}
after integrating out $\widetilde{\theta}_k$ and $\theta_k$ and taking the limit $K \rightarrow \infty$ \citep{ghahramani2005infinite}. Since we approximate the infinite IBP with a finite stick-breaking process truncated at $K^{\star}$, we can therefore approximate the IBP priors as
\begin{equation*}
    \begin{aligned}
    & \pi(\widetilde{\gamma}_{jk} \mid \widetilde{\theta}_k) \overset{\text{ind}}{\sim} \text{Bernoulli}(\widetilde{\theta}_k), ~~~ \pi(\widetilde{\theta}_k) \sim \text{Beta}\left( \frac{\widetilde{\alpha}}{K^{\star}}, 1 \right),~~~ k = 1, \ldots, K^{\star},  \\
    & \pi(\gamma_{jk} \mid \theta_k ) \overset{\text{ind}}{\sim} \text{Bernoulli}(\theta_k), ~~~ \pi(\theta_k) \sim \text{Beta}\left( \frac{\alpha}{K^{\star}}, 1 \right), ~~~ k =  1, \ldots, K^{\star}. 
    \end{aligned}
\end{equation*}
Based on these approximations, we can use the updates suggested by \citet{rovckova2016fast} and \citet{moran2019variance} for $\widetilde{\boldsymbol{\tau}} = (\widetilde{\tau}_1, \ldots, \widetilde{\tau}_K^{\star})^\top$ and $\boldsymbol{\tau} = (\tau_1, \ldots, \tau_{K^{\star}})^\top$ as
\begin{equation} \label{BiSSLB-tauk-updates}
    \widetilde{\tau}_k^{(t)} \leftarrow \frac{\tilde{\alpha}/K^{\star} + \left\|\mathbf{a}_k^{(t)}\right\|_0}{\tilde{\alpha}/K^{\star} + 1 + I} \quad \text{and} \quad
    \tau_k^{(t)} \leftarrow \frac{\alpha / K^{\star} + \left\|\mathbf{b}_k^{(t)} \right\|_0}{\alpha/K^{\star} + 1 + J}, ~~~ k = 1, \ldots, K^{\star},
\end{equation}
where $\lVert \mathbf{v} \rVert_0$ denotes the number of nonzero elements in a vector $\mathbf{v}$. 

Note that since $\widetilde{\theta}_{(k)} = \prod_{l=1}^{k} \widetilde{\nu}_{l}$ and $\theta_{(k)} = \prod_{l=1}^{k} \nu_l$ where $\widetilde{\nu}_l, \nu_l \overset{\text{iid}}{\sim} \text{Beta}(\widetilde{\alpha}, 1)$, it must also be the case that $\widetilde{\tau}_{1} > \widetilde{\tau}_{2} > \cdots > \widetilde{\tau}_{K^{\star}}$ and $\tau_{1} > \tau_{2} > \cdots > \tau_{K^{\star}}$. However, the updates \eqref{BiSSLB-tauk-updates} may not respect this implicit ordering. Thus, after updating $\{ \widetilde{\tau}_k^{(t)} \}_{k=1}^{K^\star}$ and $\{ \tau_k^{(t)} \}_{k=1}^{K^\star}$ as in \eqref{BiSSLB-tauk-updates}, we reorder the $\widetilde{\tau}_k^{(t)}$'s and the $\tau_k^{(t)}$'s in descending order to obtain our final updates for $\widetilde{\boldsymbol{\tau}}^{(t)}$ and $\boldsymbol{\tau}^{(t)}$. We also reorder the corresponding columns in $\mathbf{A}$ and $\mathbf{B}$ in the order of the original indices of the reordered $\widetilde{\boldsymbol{\tau}}^{(t)}$ and $\boldsymbol{\tau}^{(t)}$. These updates  $\widetilde{\boldsymbol{\tau}}^{(t)}$ and $\boldsymbol{\tau}^{(t)}$ are used to respectively update $a_{ik}$ and $b_{jk}$ in their respective subsequent $(t+1)$st updates \eqref{BiSSLB-refined-A-update} and \eqref{BiSSLB-refined-B-update}. Finally, we remove the zero columns and reset $K^{\star}$ to be the number of nonzero columns at the end of each $t$th iteration. 

\subsection{Initialization of BiSSLB algorithm}
Biclustering algorithms often suffer from cold start problems, and BiSSLB is no exception. Following the strategy adopted in \cite{lee2014biclustering}, we initialize the algorithm using a truncated singular value decomposition (SVD) of the observed data matrix $\mathbf{Y}$. Specifically, let
\begin{equation}
    \mathbf{Y} = \mathbf{U}\mathbf{D}\mathbf{V}^{\top},
\end{equation}
where $\mathbf{U}$ and $\mathbf{V}$ are orthogonal matrices and $\mathbf{D}$ is a diagonal matrix of the singular values of $\mathbf{Y}$.

Given an initial estimate of the number of biclusters $K$, we retain the first $K$ singular components, denoted by $\mathbf{U}_K = \mathbf{U}[:,1:K]$, $\mathbf{D}_K = \mathbf{D}[:,1:K]$, and $\mathbf{V}_K = \mathbf{V}[:,1:K]$. The latent factor matrices are then initialized as
\begin{equation}
    \mathbf{A}^{(0)} = \mathbf{U}_K \mathbf{D}_K^{-1/2} \quad \text{and}
 \quad    \mathbf{B}^{(0)} = \mathbf{V}_K \mathbf{D}_K^{-1/2}.
\end{equation}
In addition, we initialize the location parameter $\boldsymbol{\mu}$ as the zero vector. All column sparsity parameters $\widetilde{\boldsymbol{\theta}}$ and $\boldsymbol{\theta}$, are initialized to $0.5$.

When prior estimates of $\mathbf{A}$ and $\mathbf{B}$ are available, e.g. those obtained by using a different pair of spike and slab hyperparameters $(\lambda_0, \lambda_1)$, BiSSLB can alternatively be initialized using these previous MAP estimates of $\mathbf{A}$ and $\mathbf{B}$. This allows the BiSSLB algorithm to warm-start from a previously obtained solution, which can potentially achieve faster convergence for a new set of hyperparameters $(\lambda_0, \lambda_1)$ \citep{rovckova2016fast, rovckova2018spike, moran2021spike}.

\section{Performance Measures} \label{Sec:App-Performance}


Here, we describe how we obtain the performance metrics we used for evaluating biclustering performance, namely clustering error (CE), consensus score (CS), relevance, and recovery. Suppose that we have two sets of biclusters: the ground truth $\mathcal{C} = \{C_1, C_2, \ldots, C_K\}$ and the estimated biclusters $\mathcal{C}^{\prime} = \{C_1^{\prime}, C_2^{\prime}, \ldots, C_{K^{\prime}}^{\prime}\}$.  
Following \citet{patrikainen2006comparing}, several clustering evaluation measures can be derived from a confusion matrix.
\begin{enumerate}
    \item The standard confusion matrix, denoted as $\mathbf{M}^{\text{(CE)}}$, is a $K \times K^{\prime}$ matrix used to compute the CE. Each $(i,j)$th entry of $\mathbf{M}^{\text{(CE)}}$ represents the number of shared elements between clusters $i$ and $j$, i.e.,
    \begin{equation}\label{eq:Mij-CE}
        m^{\text{(CE)}}_{ij} = |C_i \cap C_j^{\prime}|.
    \end{equation}

    \item A modified confusion matrix, denoted as $\mathbf{M}^{\text{(CS)}}$, applies the Jaccard index to measure cluster similarity \citep{moran2021spike}, which is typically used to compute the CS. Each $(i,j)$th entry of $\mathbf{M}^{\text{(CS)}}$ represents the weighted number of shared elements between clusters $i$ and $j$, i.e.,
    \begin{equation}\label{eq:Mij-CS}
        m^{\text{(CS)}}_{ij} = J(C_i, C_j^{\prime}) = \frac{|C_i \cap C_j^{\prime}|}{|C_i \cup C_j^{\prime}|},
    \end{equation}
    where $J(C_i, C_j^{\prime})$ denotes the Jaccard index.
\end{enumerate}
To compute the CE and CS, we first construct the confusion matrices $\mathbf{M}^{\text{(CE)}}$ and $\mathbf{M}^{\text{(CS)}}$ in \eqref{eq:Mij-CE} and \eqref{eq:Mij-CS} respectively. We then apply the Hungarian algorithm \citep{munkres1957algorithms} to determine the optimal one-to-one assignment between $\mathcal{C}$ and $\mathcal{C}^{\prime}$. The final CE or CS is calculated by summing the matched entries and normalizing by the maximum number of biclusters:
\begin{equation}
    \text{Score} = \frac{\sum_{(i, j) \in \text{optimal match}} m_{ij}}{\max\{|\mathcal{C}|, |\mathcal{C}^{\prime}|\}}.
\end{equation}
As noted by \citet{li2020bayesian}, the CE accounts for the different bicluster sizes via set cardinality (intersection count). In contrast, by using the Jaccard index, CS gives equal weight to all bicluster pairs regardless of size.

Following \citet{moran2021spike}, we also define the relevance and recovery scores as
\begin{equation}
    \begin{aligned}
        \text{Relevance} &= \frac{1}{|\mathcal{C}^{\prime}|} \sum_{C_j^{\prime} \in \mathcal{C}^{\prime}} \max_{C_i \in \mathcal{C}} J(C_i, C_j^{\prime}), \\
        \text{Recovery}  &= \frac{1}{|\mathcal{C}|} \sum_{C_i \in \mathcal{C}} \max_{C_j^{\prime} \in \mathcal{C}^{\prime}} J(C_i, C_j^{\prime}).
    \end{aligned}
\end{equation}
Relevance measures how similar on average the estimated biclusters are to the true biclusters, while recovery measures how similar the true biclusters are to the estimated biclusters on average \citep{moran2021spike}. The recovery score is similar to CS, but CS penalizes overestimation of the true number of biclusters (i.e., estimating too many biclusters results in a lower CS) \citep{moran2021spike}.

\vskip 0.2in
\bibliography{paper}
\newpage

\end{document}